\documentclass{article}

\usepackage[labelfont=bf]{caption} 
\usepackage[font={small}]{caption}   

\usepackage{graphicx}
\usepackage{dcolumn}
\usepackage{bm}
\usepackage[pdftex,colorlinks=true,linkcolor=blue,citecolor=blue]{hyperref}
\usepackage[utf8]{inputenc}
\usepackage{lipsum}  
\usepackage{amsmath}
\usepackage{authblk}
\usepackage{amsfonts}
\usepackage{placeins}
\usepackage[numbers, sort&compress]{natbib}  
\usepackage{bibunits}

\usepackage{amssymb}
\usepackage{pdflscape} 

    \usepackage[table, usenames, dvipsnames]{xcolor}
    \usepackage{makecell}
    \usepackage{boldline}
    \setcellgapes{3pt}

\usepackage{textgreek}      

\usepackage[strict]{changepage} 
\usepackage{nicefrac} 

\defaultbibliographystyle{naturemag}
\defaultbibliography{thesisrefsol}

\title{Electrically driven linear optical isolation through phonon mediated Autler-Townes splitting}

\author{
    Donggyu B. Sohn $^\dag$, Oğulcan E. Örsel $^\dag$, Gaurav Bahl \\
    Department of Mechanical Science and Engineering, \\
    University of Illinois at Urbana–Champaign, Urbana, IL 61801 USA \\
    $^\dag$ These authors contributed equally
}

\date{}

\begin{document}
\begin{bibunit}

\maketitle


\begin{abstract}

Optical isolators are indispensible components in nearly all photonic systems as they help ensure unidirectionality and provide crucial protection from undesirable reflections. While commercial isolators are exclusively built on magneto-optic (MO) principles they are not readily implemented within photonic integrated circuits due to the need for specialized materials. Importantly, the MO effect is generally weak, especially at shorter wavelengths.
These challenges as a whole have motivated extensive research on non-MO alternatives~\cite{Hwang:97,KangM.2011,Doerr:11, lira2012,Tzuang2014, Li:2014vo,Sounas:14,Dong2015,Kim2015,Kim2016,Ruesink:16, Fang:2017,Sohn18,Kittlaus2018,Peterson:18,Shi:2018aa,Sohn:2019aa,Tian:2020ti,sarabalis2020,Dostart:2021uh,Kim:2021te,Kittlaus:2021wq}. 
To date, however, no alternative technology has managed to simultaneously combine linearity (i.e. no frequency shift), linear response (i.e. input-output scaling), ultralow insertion loss, and large directional contrast on-chip.
Here we demonstrate an optical isolator design that leverages the unbeatable transparency of a short, high quality dielectric waveguide, with the near-perfect attenuation from a critically-coupled absorber.
Our design concept is implemented using a lithium niobate racetrack resonator in which phonon mediated~\cite{Sohn18} Autler-Townes splitting (ATS)~\cite{Peng:2014ui,Kim2016,Shi:2018aa,Zhang:2019ww} breaks the chiral symmetry of the resonant modes. 
We demonstrate on-chip optical isolators at wavelengths one octave apart near 1550 nm and 780 nm, fabricated from the same lithium niobate-on-insulator wafer. Linear optical isolation is demonstrated with simultaneously $<$1 dB insertion loss, $>$39 dB contrast, and bandwidth as wide as the optical mode that is used. 
Our results outperform the current best-in-class MO isolator on-chip on both insertion loss and isolator figures-of-merit, and demonstrate a lithographically defined wavelength adaptability that cannot yet be achieved with any MO isolator.

\end{abstract}

\maketitle

Magneto-optic (MO) isolators, based on asymmetric Faraday rotation, have long dominated nonreciprocal photonic device technologies as they simultaneously provide low insertion loss, high directional contrast, and wide bandwidth. 
While a number of attempts have been made to bring these devices on chip~\cite{Ross:11,Ghosh:12aa,Huang:17,Zhang:2017wq,Du:2018wy,Zhang:19,Yan:20}, with some measure of success, there remain many unresolved challenges that have held back full adoption.
The foremost among these is the lack of appropriate materials in photonics foundries, and the additional technical constraint of magnetic biasing.
Recent results~\cite{Yan:20,Zhang:19,Zhang:2017wq} on integrated magneto-optic isolators have been successful at achieving low insertion loss over modest bandwidth, however they cannot easily be adapted to any wavelength of choice since the Faraday effect is chromatic, i.e. is a strong function of the wavelength.
Moreover, the MO material stack is intrinsically lossy and therefore the best current examples~\cite{Yan:20,Zhang:19,Zhang:2017wq} attempt to use the minimum possible amount of MO material.

Due to a wide recognition of these limitations, extensive research has been performed on alternative isolator and circulator technologies that leverage 
synthetic fields~\cite{Tzuang2014, Li:2014vo, Fang:2017,Kim:2021te}, 
optomechanics~\cite{KangM.2011,Sounas:14,Kim2015,Dong2015,Kim2016,Ruesink:16},
acousto-optics~\cite{Sohn18,Sohn:2019aa,Kittlaus2018,Kittlaus:2021wq,Hwang:97,Tian:2020ti,sarabalis2020}, electro-optics~\cite{Doerr:11,lira2012,Shi:2018aa,Dostart:2021uh}, spinning resonators~\cite{Maayani:2018vc}, and chirally pumped atoms~\cite{Scheucher:16}. 
These techniques may be generally considered as variations on spatio-temporal modulation and use some momentum conservation rule or momentum bias to break reciprocity. Often these approaches operate over a narrow bandwidth, which is quite acceptable for a wide variety of single-frequency laser applications, for instance in ultra-stable sources~\cite{Spencer:2018ta, Lucas:2020tz}, LIDAR~\cite{Poulton:2017tn}, frequency combs~\cite{DelHaye:2007tv}, and atomic referencing~\cite{Hummon:2018wp, Knappe:04, Newman:2019wo}. A number of these alternative techniques have demonstrated very large optical contrast~\cite{Dostart:2021uh} (see also Supplement \S\ref{sec:LitReview}),
but ultimately the ability to provide low insertion loss remains a huge technical challenge. 

\begin{figure}[th!]
    \begin{adjustwidth}{-1in}{-1in}
    \centering
    \includegraphics[width=1.1\textwidth]{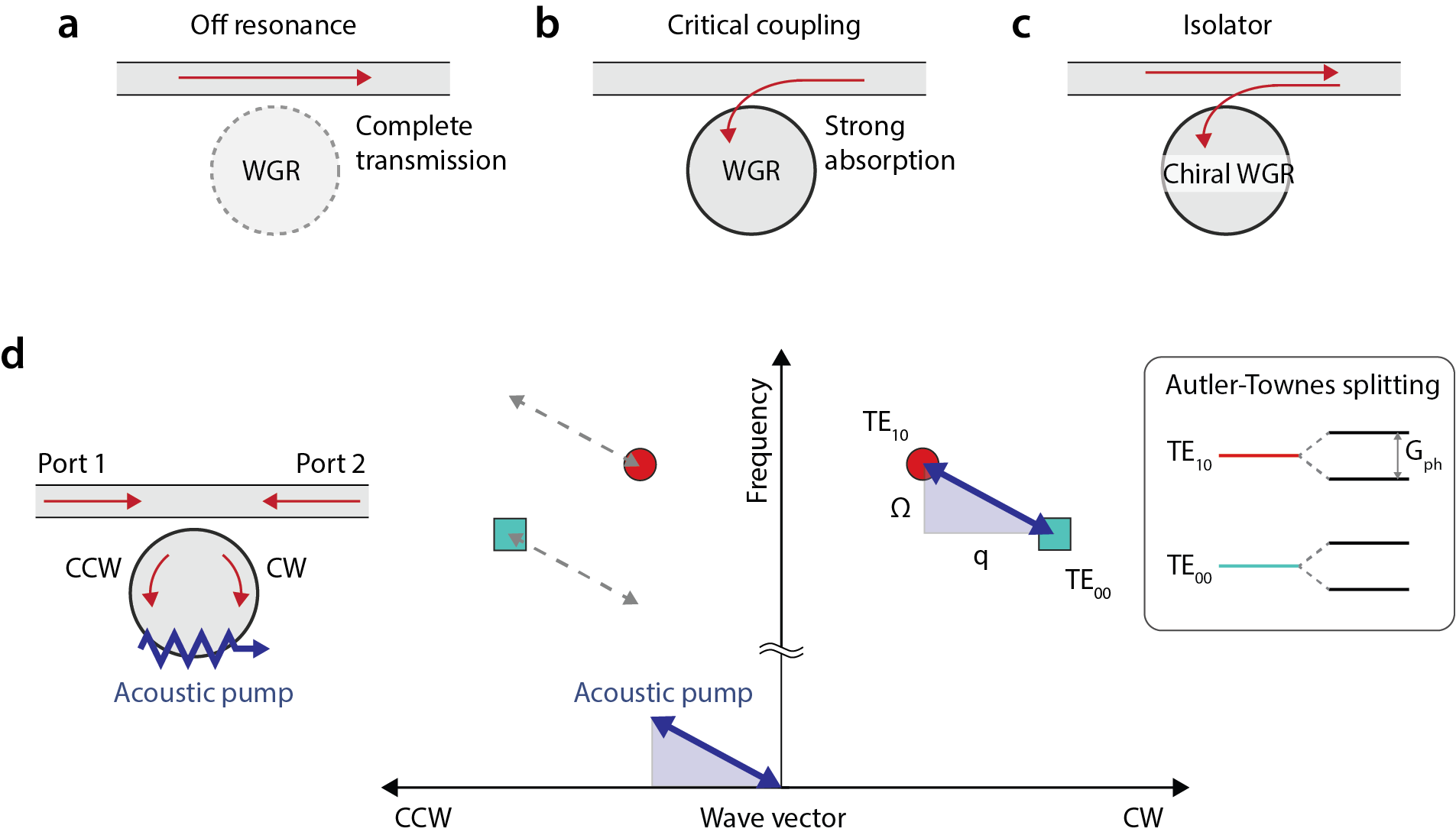}
    \caption{
        \textbf{Optical isolation with a chiral absorber.}
        \textbf{(a)} For a waveguide-resonator system, any light that propagates detuned far from the resonance of the absorber (here, a whispering gallery resonator or WGR) passes through the waveguide uninterrupted. Propagation loss is almost entirely dictated by the waveguide quality.
        \textbf{(b)} If the waveguide and WGR are critically coupled, a nearly complete on-resonance absorption situation is achieved.
        \textbf{(c)} For a photonic WGR with chiral density of states, there exist frequencies where depending on propagation direction in the waveguide, complete transparency and strong attenuation can be simultaneously achieved.
        \textbf{(d)} We use phonon-mediated Autler-Townes splitting (ATS) in a two-level photonic atom (the TE$_{10}$ and TE$_{00}$ modes of the WGR act as the coupled levels) to produce the chiral WGR. In the diagram shown, the ATS appears only for optical wavevectors corresponding to CW circulation due to the acousto-optic phase matching condition. The phonon-enhanced optomechanical coupling rate $G_{ph}$ determines the amount of level splitting that is generated.
        CW = Clockwise. CCW = Counter-clockwise.
    }
    \label{fig:1}
    \end{adjustwidth}
\end{figure}

The absolute ideal for a two-port low loss device on chip is simply a high quality linear waveguide of short length. This represents the best case for achieving the lowest forward insertion loss in the `high transparency' direction of an isolator.     %
We now introduce a narrow-band absorber, e.g. a high-Q resonator, that is shunt-coupled to this ideal waveguide~\cite{Peterson:18,Kim2016}. If this absorber is detuned more than a few linewidths from the frequency of interest, it will not be accessible to light propagating within the waveguide and a high transparency is observed (Fig.~\ref{fig:1}a). On the other hand, when light is on resonance with the absorber, and if the absorber is critically coupled to the waveguide (Fig.~\ref{fig:1}b), then a giant attenuation factor can be achieved. Here, critical coupling occurs when the intrinsic loss rate for the WGR mode is matched to the coupling rate from the external interface (Supplement~\S\ref{sup:critical}).
It is this combination of near-ideal transparency and giant attenuation that we wish to simultaneously exploit \cite{Kim2016,Shi:2018aa}, and it can be achieved using a narrow-band absorber with broken chiral symmetry (Fig.~\ref{fig:1}c). 

\vspace{12pt}

In this Letter we report on a method to induce very large chiral asymmetry in a two-level photonic atom using phonon mediated Autler-Townes splitting (ATS)~\cite{Kim2016,Shi:2018aa,Sohn18}.
When coupled with a waveguide, the resulting isolator exhibits near ideal characteristics. The two-level photonic atom is produced using a whispering gallery racetrack resonator (WGR) that supports two families of optical modes belonging to TE$_{10}$ and TE$_{00}$ families.
We identify a mode pair that is closely spaced in frequency near a mode family crossing, with the TE$_{00}$ mode ($\omega_1, k_1$) and TE$_{10}$ mode ($\omega_2, k_2$) having distinct frequencies and wavevectors as shown in Fig.~\ref{fig:1}d.
These two modes can be unidirectionally coupled through acousto-optic scattering, as long as the difference of frequency ($\Omega = \omega_2 - \omega_1$) and momentum ($q = k_2 - k_1$) between the optical modes matches the frequency $\Omega$ and momentum $q$ of an acoustic excitation of the material. In addition, the overlap integral between the acoustic mode and optical modes must be non-zero~\cite{Sohn18}.
Both these requirements, while tricky, can be simultaneously achieved by the engineering of a 2D `texture' for the acoustic excitation of the resonator~\cite{Sohn18,Sohn:2019aa}, such that there is a non-zero momentum along the WGR circuit, while a standing wave exists in the transverse direction.
When the phonon-enhanced optomechanical coupling rate ($G_{ph}$) exceeds the optical loss rates ($G_{ph}>\sqrt{\kappa_{1}\kappa_{2}}$) we enter the strong coupling or ATS regime (see Supplement \S\ref{sec:ATSTheory}). This regime manifests as a unidirectional hybridization of the selected TE$_{10}$ and TE$_{00}$ modes, modifying their dispersion and frequency-splitting the resonant absorption in the phase matched direction only. In this analogy, the $G_{ph}$ determines the amount of frequency splitting and is equivalent to the Rabi frequency in atomic ATS.
In this situation, light propagating at the original optical resonance frequencies ($\omega_1$ and $\omega_2$) in the waveguide no longer interacts with the resonator and simply propagates through. This will be the case as long as the process is reasonably well phase matched, and the coupling rate $G_{ph}$ is sufficiently large.
In the non-phase matched direction, however, the original optical modes remain unmodified. As a result, if we can design the system to achieve critical coupling for the TE$_{10}$ mode, then giant contrast is simultaneously achieved (see Supplement \S\ref{sup:critical}).
As a result of this architecture our approach achieves unidirectional strong coupling and large nonreciprocity with a single RF input, without requiring multiple phase-shifted signals, which is common in other spatiotemporal modulation approaches.

\vspace{12pt}

\begin{figure}[tp]
    \begin{adjustwidth}{-1in}{-1in}
    \centering
    \includegraphics[width=1.1\textwidth]{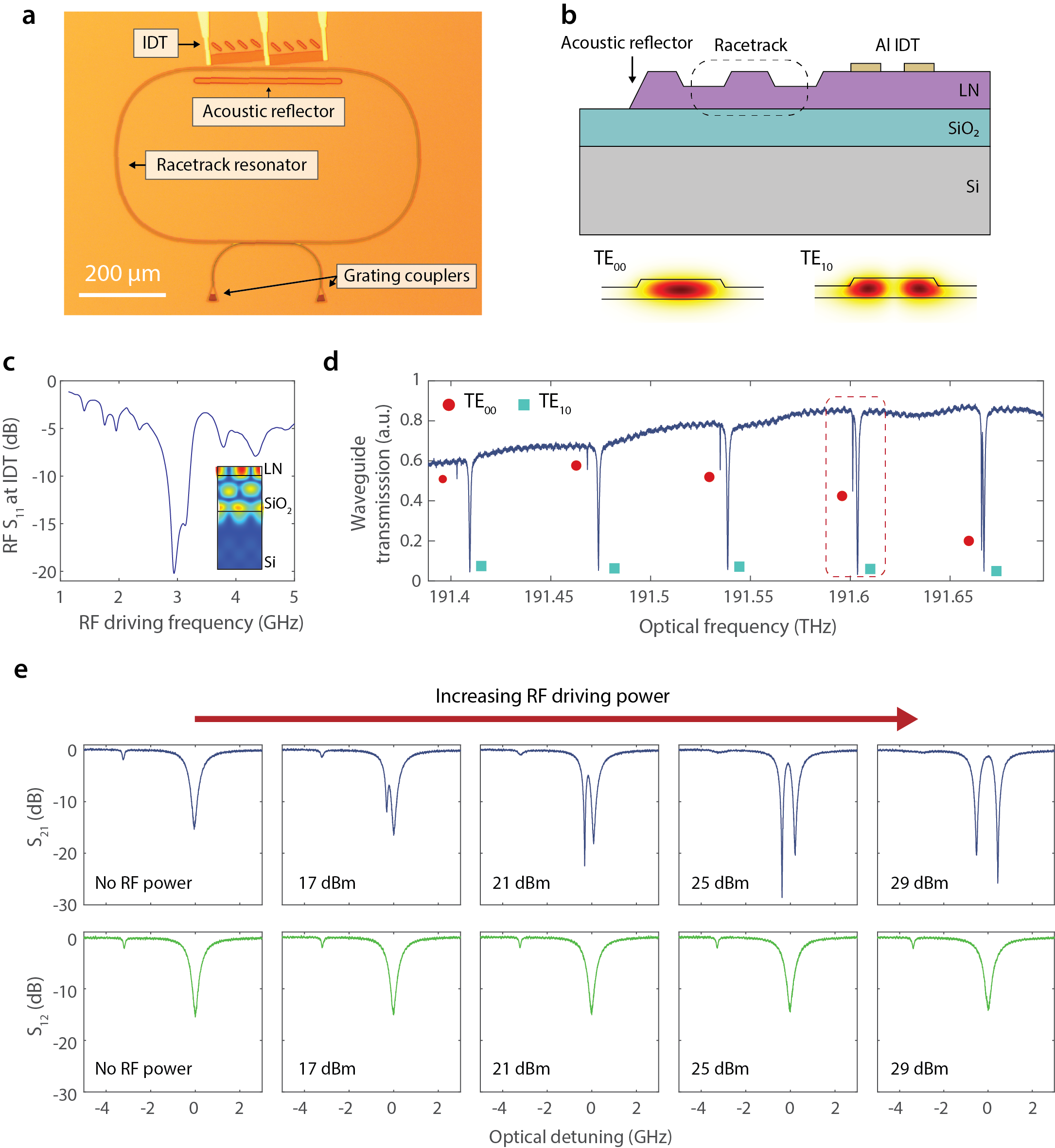}
    \caption{
        \textbf{Phonon-mediated ATS isolator implementation and characterization.} 
        \textbf{(a)} Microscope image of the lithium niobate optical isolator. The device is composed of a racetrack WGR with adjacent waveguide. The required 2D surface acoustic wave `texture' \cite{Sohn18, Sohn:2019aa} is produced using an angled interdigitated actuator (IDT) and acoustic reflector on opposite sides of the racetrack.
        \textbf{(b)} Cross-section of the acousto-optic interaction region shows the racetrack resonator (modes are presented in inset) with the aluminum IDT and acoustic reflector. The device is unreleased.
        \textbf{(c)} IDT characterization using RF reflection measurement (RF $S_{11}$ parameter) shows that the surface acoustic wave in this device is efficiently generated around 3 GHz, as determined by the designed IDT pitch and the surface acoustic wave speed.
        \textbf{(d)} Optical characterization (transmission through waveguide) without any applied RF stimulus.
        \textbf{(e)} As we increase the RF power applied to the IDT, the optical $S_{21}$ (transmission from port 1 to port 2) and $S_{12}$ (transmission from port 2 to port 1) measurements performed through the waveguide exhibit phonon mediated ATS in only one direction.
        A closer examination of this isolator is presented in Fig.~\ref{fig:3}a.
    }
    \label{fig:2}
    \end{adjustwidth}
\end{figure}

For experimental implementation, we used a lithium niobate-on-insulator (LNOI) integrated photonics platform. 
The wide bandgap of lithium niobate (LN) imparts a broad transparency window spanning 250 nm to 5300 nm.
Additionally, the high piezoelectric coefficient in LN allows very efficient actuation of surface acoustic waves via RF stimulus~\cite{Gong:13}, which is a specific advantage for this device. In contrast to previous efforts that used aluminum nitride as the photonic and piezoelectric material~\cite{Sohn:2019aa,Sohn18}, here we are able achieve much higher optical Q-factors ($10^7$ at 780 nm, $3.5\times 10^6$ at 1550 nm) and significantly better electromechanical transduction efficiency (about 10x higher at $40\%$).
The two-mode racetrack resonators and adjacent single-mode waveguides (Fig.~\ref{fig:2}a) are etched with ridge waveguide geometry (see Supplement \S\ref{sec:Fabrication}). 
Grating couplers covering the wavelength range of interest are fabricated on either end of the waveguide to provide off-chip optical access to the isolator.
Finally, to ensure the transverse standing wave characteristic for the acoustic excitation, we fabricate an acoustic reflector by fully etching the LN thin film on the far side of the racetrack (Fig.~\ref{fig:2}a,b).

We first characterize the primary acoustic and optical components of the isolator.
Measurement of the RF reflection coefficient (S$_{11}$) of the IDT shows a dip near 3 GHz, confirming excitation of the surface acoustic wave (Fig.~\ref{fig:2}c).
The optical states of the racetrack resonator are measured by probing transmission through the adjacent waveguide. 
As seen in Fig.~\ref{fig:2}d, two optical mode families can be identified, with the TE$_{10}$ family better coupled to the waveguide due to its larger evanescent field. In fact, the TE$_{10}$ mode is critically coupled to the waveguide to ensure maximum attenuation in the backward direction. The TE$_{00}$ mode is kept intentionally dark (i.e. under-coupled) since it is helpful to both lower the $G_{ph}$ requirement and, as we discuss later, to suppress sideband generation.
The device is designed such that there is a mode crossing near the wavelength of interest.
For the specific device shown, we find an optical mode pair located near 192.6 THz (1556 nm) that has a frequency separation of $\sim 3$ GHz, which is similar to the acoustic frequency. 
Finally, a constant RF tone at 3 GHz is applied to the IDT, which launches the 2D surface acoustic wave and hybridizes the two optical modes producing the ATS.

\begin{figure}[tp!]
    \begin{adjustwidth}{-1in}{-1in}
    \centering
    \includegraphics[width=1\textwidth]{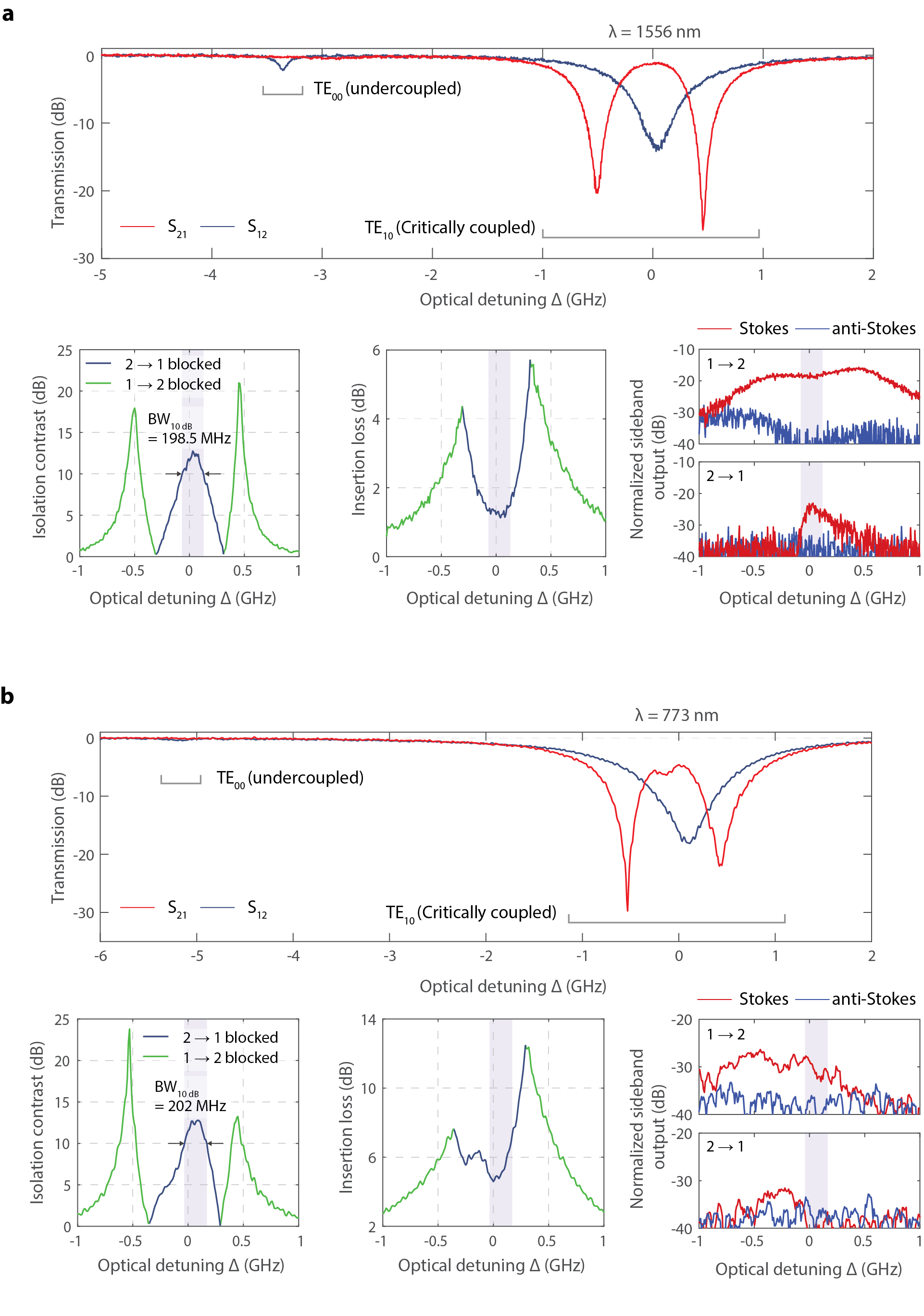}
    \caption{
        \textbf{Experimental demonstration of phonon mediated ATS isolators near 1550 and 780 nm.}
        Detuning $\Delta$ is defined relative to the unperturbed TE$_{10}$ mode.
        \textbf{(a)} Measurement results from a 1556 nm device for 3 GHz applied RF with 29 dBm power (measured $G_{ph} \approx 0.98$ GHz).
        \textbf{(b)} Measurement results from a 773 nm device for 5.05 GHz applied RF with 25 dBm power (measured $G_{ph} \approx 0.99$ GHz).
        A discussion is provided in the main text and additional details on the device parameters can be found in Supplementary Table \ref{tab:DeviceParameters}.
    }
    \label{fig:3}
    \end{adjustwidth}
\end{figure}

We experimentally measure the non-reciprocal transmission using optical heterodyne detection (see Supplement \S\ref{sec:Setup}) which enables separate measurement of the carrier signal transmission and its sidebands.
Fig.~\ref{fig:2}e shows the evolution of the optical spectrum for increasing RF driving power, in both forward and backward directions through the waveguide ports. 
We define the frequency of the original TE$_{10}$ mode as the zero detuning $\Delta = 0$ point.
In the phase matched direction (forward), the optical modes experience a clear ATS phenomenon, with the central transmission at zero detuning approaching unity for large drive power. At the same time, the optical modes remain unperturbed in the non-phase matched direction (backward).

A closer examination of the telecom (1550 nm) wavelength isolator is presented in Fig.~\ref{fig:3}a. 
We observe that this device simultaneously achieves 1.13 dB forward insertion loss with 12.75 dB peak contrast using 29 dBm of RF driving power. This drive level corresponds to a phonon-enhanced optomechanical coupling rate $G_{ph}$ of 0.98 GHz, and we are currently prevented from increasing this further due to the power handling capability of the IDTs. 
The nonreciprocal contrast here is limited only by the degree to which we are able to approach the critical coupling in the fabricated device.
While our main attention is on the central isolation behavior, we also observe appreciable nonreciprocity on the the wings corresponding to the dressed states, although here we must reverse the description of forward and backward propagation (as those are arbitrary definitions). 
During the experiment, we are also able to monitor the $\pm 3$ GHz sidebands of the carrier signal, and we find that they are consistently $\sim 20$ dB below (or more) the original carrier signal power. 
This is thanks to a combination of the very high $G_{ph}$ which significantly reduces interaction between the waveguide and WGR at $\Delta = 0$, and also thanks to the undercoupled TE$_{00}$ mode which does not couple well to the waveguide. Further discussion is provided in the Supplement \S\ref{sec:Sidebands}.

Since all features of this device are lithographically defined, and do not depend on any wavelength-dependent material gyrotropic property (e.g. the Faraday rotation effect in magneto-optics), we are able to shift the operational wavelength of this isolator freely within the transparency range of lithium niobate. In fact, the optomechanical coupling $G_{ph}$ is directly proportional to the optical frequency, implying that isolators operating at short wavelengths with comparable parameters will require lower power. There is, however, the possibility that this advantage is offset by the generally higher surface scattering induced loss at short wavelengths. In this context, we developed a 780 nm demonstration (Fig.~\ref{fig:3}b) for the isolator using the same LNOI substrate. This device simultaneously achieves 4.76 dB forward insertion loss with 12.86 dB peak contrast, using 25 dBm of RF driving power to reach $G_{ph} \approx 0.99$ GHz. While the insertion loss achieved in this device was not as low, the sideband generation is significantly reduced since the TE$_{00}$ mode is almost completely dark. 

\vspace{12pt}

\begin{figure}[th]
    \begin{adjustwidth}{-1in}{-1in}
    \centering
    \includegraphics[width=1\textwidth]{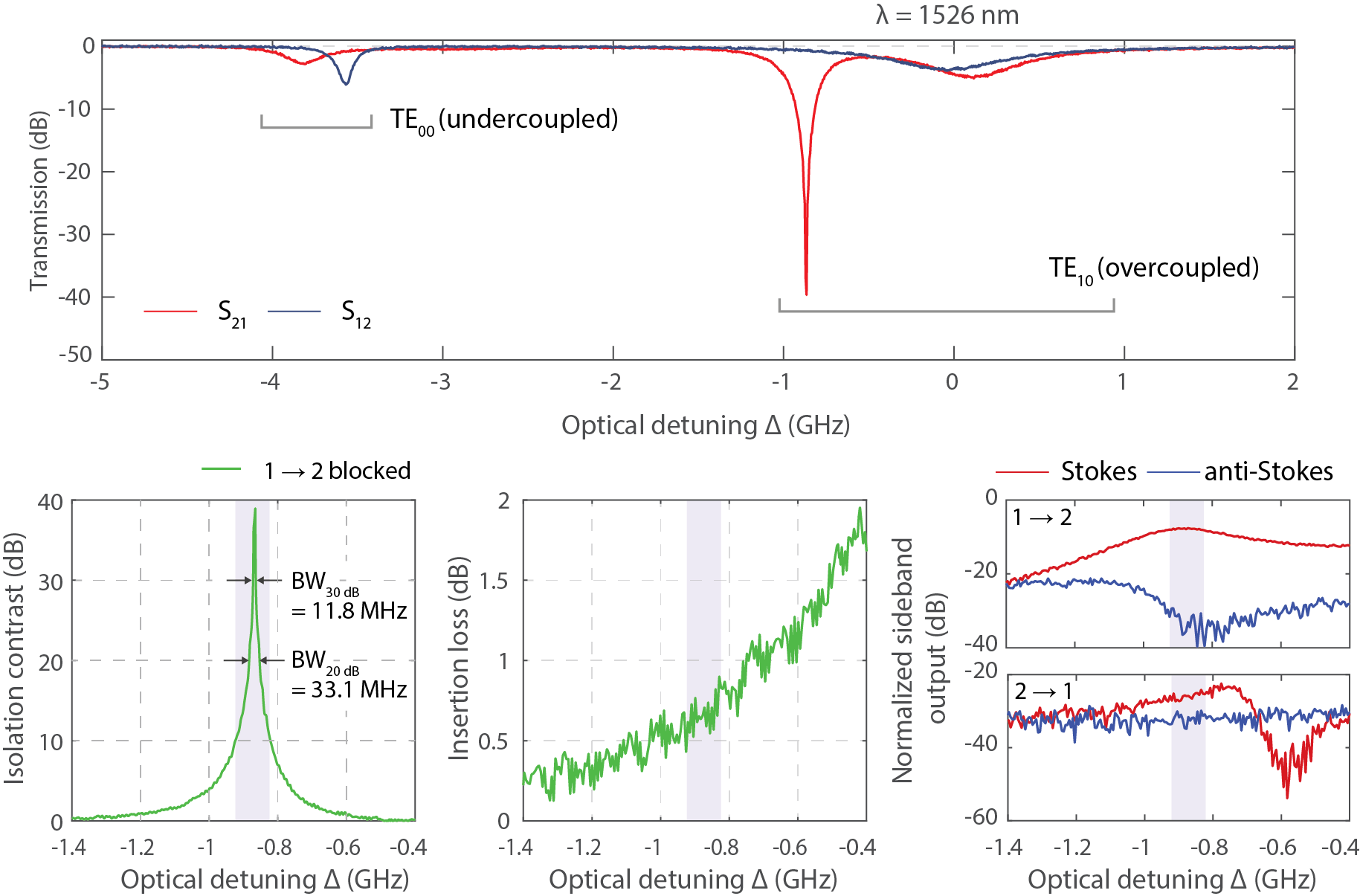}
    \caption{
        \textbf{Demonstration of giant isolation using dressed state nonreciprocity.}
        Detuning $\Delta$ is defined relative to the unperturbed TE$_{10}$ mode. 
        Measurement results from a 1526 nm isolator for 3.04 GHz applied RF with 29 dBm power (measured $G_{ph} \approx 0.76$ GHz).
        Another example of this effect for a device operating near 780 nm is shown in Supplementary Fig.~\ref{sup:asym780}.
        Additional details on the device parameters can be found in Supplementary Table \ref{tab:DeviceParameters}. 
    }
    \label{fig:4}
    \end{adjustwidth}
\end{figure}

In the above ATS-based approach, the isolation contrast is primarily dictated by the criticality of coupling between the waveguide and the optical mode (here, the TE$_{10}$ mode) and is therefore strongly dependent on the fabricated geometry. Contrast factors greater than 25 dB can therefore be difficult to achieve because of the matching requirement with the intrinsic resonator loss.
One solution to the contrast problem is to instead leverage the nonreciprocal wings to either side of the central nonreciprocal band. In these bands, the contrast is dictated by the placement of, and coupling to the dressed states. The contrast is therefore a function of $G_{ph}$, and of the relative mismatch $\delta = \omega_1 - \omega_2 + \Omega$ between the optical mode separation and the applied phonon frequency. 
Additional discussion of this situation is provided in the Supplement \S\ref{sec:WingOptimization}.
As a key result, we find that if the TE$_{10}$ mode is intentionally overcoupled, and the mode mismatch is $\delta < 0$, then there exists a choice of $G_{ph}$ where critical coupling is achievable near the lower dressed state. Conversely, for $\delta > 0$ the critical coupling is achievable near the upper dressed state.
A dramatic example of this situation is presented for an overcoupled 1538 nm isolator in Fig.~\ref{fig:4}, where on the lower dressed state we can simultaneously observe $0.65$ dB insertion loss with $39.3$ dB contrast, using 29 dBm of RF driving power to reach $G_{ph} \approx 0.76$ GHz. 
Here, the spacing between the optical modes is 3.72 GHz while the RF drive is at 3 GHz, implying a $\delta \approx -0.72$ GHz.
We also implemented this effect with opposite $\delta$ using a device operating near 780 nm (see Supplementary Fig.~\ref{sup:asym780}).    

We can understand this phenomenon fairly intuitively by considering instead a sideband picture, which works since the mismatch $\delta$ puts the system in a regime between pure ATS and the ac Stark effect (the latter explains the noticeable frequency shifts of the modes).
As we scan the overcoupled TE$_{10}$ mode (i.e. intrinsic loss rate $<$ extrinsic coupling rate), we find a range of frequencies where the scattering into the nearly dark TE$_{10}$ is resonantly enhanced. This implies that as $G_{ph}$ increases, the effective intrinsic loss rate of the TE$_{10}$ mode over this band increases, and can approach the extrinsic coupling rate to hit criticality. This intuitive explanation is also experimentally supported by the increased Stokes sideband power in the $2 \rightarrow 1$ case in Fig.~\ref{fig:4} which indicates increased scattering into the TE$_{00}$ mode.
While this approach is much more sensitive to the value of $G_{ph}$, unlike the central isolation band, the tunability of this operation regime can be advantageous and ultimately delivers better results in the devices shown in this paper.

Since isolators are often used in cascaded configurations, citing insertion loss (IL) or isolation contrast (IC) alone is not meaningful. Instead, as discussed previously~\cite{Kim2016}, the ratio of IC per dB of IL is the relevant figure of merit (FoM) that accounts for cascading. We find that this device (Fig.~\ref{fig:4}) not only provides IC at par with the best magneto-optic isolators, but significantly outperforms them on IL and the isolator FoM (comparison in Supplement \S\ref{sec:LitReview}). The main disadvantage of our approach compared to magneto-optics is the narrower operational bandwidth, but this could be addressed in the future with designs that achieve higher $G_{ph}$.

\vspace{12pt}

The lack of optical isolators in photonic integrated circuits has long been a technological hurdle, and their absence is uniquely felt by contemporary quantum and atomic microsystems.
Isolator and circulator banks are extensively used in low-temperature quantum technologies to suppress thermal noise, but stray magnetic fields from available isolators can be problematic for superconducting materials, e.g. in qubits. 
Magnetless isolators at visible and near-IR wavelengths are also critically needed to enable chip scale atom-photon integration~\cite{Knappe:04, Hummon:2018wp, Riedel:2010te}, as they can circumvent undesirable Zeeman shifts \cite{Blanshan:05}.
The isolators that we have demonstrated are extremely well-positioned to address these applications, as they exhibit narrowband performance at par with current off-chip magneto-optic isolators, while at the same time providing access to wavelength ranges that are challenging with magneto-optic techniques.


\vspace{24pt}

\vspace{24pt}

{\footnotesize \putbib}

{\footnotesize \bibliography{thesisrefsol}}

\begin{thebibliography}{10}
\expandafter\ifx\csname url\endcsname\relax
  \def\url#1{\texttt{#1}}\fi
\expandafter\ifx\csname urlprefix\endcsname\relax\def\urlprefix{URL }\fi
\providecommand{\bibinfo}[2]{#2}
\providecommand{\eprint}[2][]{\url{#2}}

\bibitem{Hwang:97}
\bibinfo{author}{Hwang, I.~K.}, \bibinfo{author}{Yun, S.~H.} \&
  \bibinfo{author}{Kim, B.~Y.}
\newblock \bibinfo{title}{All-fiber-optic nonreciprocal modulator}.
\newblock \emph{\bibinfo{journal}{Opt. Lett.}} \textbf{\bibinfo{volume}{22}},
  \bibinfo{pages}{507--509} (\bibinfo{year}{1997}).

\bibitem{KangM.2011}
\bibinfo{author}{Kang, M.~S.}, \bibinfo{author}{Butsch, A.} \&
  \bibinfo{author}{Russell, P. S.~J.}
\newblock \bibinfo{title}{Reconfigurable light-driven opto-acoustic isolators
  in photonic crystal fibre}.
\newblock \emph{\bibinfo{journal}{Nat. Photonics}}
  \textbf{\bibinfo{volume}{5}}, \bibinfo{pages}{549--553}
  (\bibinfo{year}{2011}).

\bibitem{Doerr:11}
\bibinfo{author}{Doerr, C.~R.}, \bibinfo{author}{Dupuis, N.} \&
  \bibinfo{author}{Zhang, L.}
\newblock \bibinfo{title}{Optical isolator using two tandem phase modulators}.
\newblock \emph{\bibinfo{journal}{Opt. Lett.}} \textbf{\bibinfo{volume}{36}},
  \bibinfo{pages}{4293--4295} (\bibinfo{year}{2011}).

\bibitem{lira2012}
\bibinfo{author}{Lira, H.}, \bibinfo{author}{Yu, Z.}, \bibinfo{author}{Fan, S.}
  \& \bibinfo{author}{Lipson, M.}
\newblock \bibinfo{title}{Electrically driven nonreciprocity induced by
  interband photonic transition on a silicon chip}.
\newblock \emph{\bibinfo{journal}{Phys. Rev. Lett.}}
  \textbf{\bibinfo{volume}{109}}, \bibinfo{pages}{033901}
  (\bibinfo{year}{2012}).

\bibitem{Tzuang2014}
\bibinfo{author}{Tzuang, L.~D.}, \bibinfo{author}{Fang, K.},
  \bibinfo{author}{Nussenzveig, P.}, \bibinfo{author}{Fan, S.} \&
  \bibinfo{author}{Lipson, M.}
\newblock \bibinfo{title}{Non-reciprocal phase shift induced by an effective
  magnetic flux for light}.
\newblock \emph{\bibinfo{journal}{Nat. Photonics}}
  \textbf{\bibinfo{volume}{8}}, \bibinfo{pages}{701} (\bibinfo{year}{2014}).

\bibitem{Li:2014vo}
\bibinfo{author}{Li, E.}, \bibinfo{author}{Eggleton, B.~J.},
  \bibinfo{author}{Fang, K.} \& \bibinfo{author}{Fan, S.}
\newblock \bibinfo{title}{Photonic aharonov--bohm effect in photon--phonon
  interactions}.
\newblock \emph{\bibinfo{journal}{Nature Communications}}
  \textbf{\bibinfo{volume}{5}}, \bibinfo{pages}{3225} (\bibinfo{year}{2014}).

\bibitem{Sounas:14}
\bibinfo{author}{Sounas, D.~L.} \& \bibinfo{author}{Al{\`u}, A.}
\newblock \bibinfo{title}{Angular-momentum-biased nanorings to realize
  magnetic-free integrated optical isolation}.
\newblock \emph{\bibinfo{journal}{ACS Photonics}} \textbf{\bibinfo{volume}{1}},
  \bibinfo{pages}{198--204} (\bibinfo{year}{2014}).

\bibitem{Dong2015}
\bibinfo{author}{Dong, C.-H.} \emph{et~al.}
\newblock \bibinfo{title}{Brillouin-scattering-induced transparency and
  non-reciprocal light storage}.
\newblock \emph{\bibinfo{journal}{Nat. Commun.}} \textbf{\bibinfo{volume}{6}}
  (\bibinfo{year}{2015}).

\bibitem{Kim2015}
\bibinfo{author}{Kim, J.}, \bibinfo{author}{Kuzyk, M.~C.},
  \bibinfo{author}{Han, K.}, \bibinfo{author}{Wang, H.} \&
  \bibinfo{author}{Bahl, G.}
\newblock \bibinfo{title}{{Non-reciprocal Brillouin scattering induced
  transparency}}.
\newblock \emph{\bibinfo{journal}{Nat. Phys}} \textbf{\bibinfo{volume}{11}},
  \bibinfo{pages}{275--280} (\bibinfo{year}{2015}).

\bibitem{Kim2016}
\bibinfo{author}{Kim, J.}, \bibinfo{author}{Kim, S.} \& \bibinfo{author}{Bahl,
  G.}
\newblock \bibinfo{title}{Complete linear optical isolation at the microscale
  with ultralow loss}.
\newblock \emph{\bibinfo{journal}{Sci. Rep.}} \textbf{\bibinfo{volume}{7}},
  \bibinfo{pages}{1647} (\bibinfo{year}{2017}).

\bibitem{Ruesink:16}
\bibinfo{author}{Ruesink, F.}, \bibinfo{author}{Miri, M.-A.},
  \bibinfo{author}{Al{\`u}, A.} \& \bibinfo{author}{Verhagen, E.}
\newblock \bibinfo{title}{Nonreciprocity and magnetic-free isolation based on
  optomechanical interactions}.
\newblock \emph{\bibinfo{journal}{Nat. Commun.}} \textbf{\bibinfo{volume}{7}},
  \bibinfo{pages}{13662} (\bibinfo{year}{2016}).

\bibitem{Fang:2017}
\bibinfo{author}{Fang, K.} \emph{et~al.}
\newblock \bibinfo{title}{Generalized non-reciprocity in an optomechanical
  circuit via synthetic magnetism and reservoir engineering}.
\newblock \emph{\bibinfo{journal}{Nat. Phys}} \textbf{\bibinfo{volume}{13}},
  \bibinfo{pages}{465--471} (\bibinfo{year}{2017}).

\bibitem{Sohn18}
\bibinfo{author}{Sohn, D.~B.}, \bibinfo{author}{Kim, S.} \&
  \bibinfo{author}{Bahl, G.}
\newblock \bibinfo{title}{Time-reversal symmetry breaking with acoustic pumping
  of nanophotonic circuits}.
\newblock \emph{\bibinfo{journal}{Nat. Photonics}}
  \textbf{\bibinfo{volume}{12}}, \bibinfo{pages}{91--97}
  (\bibinfo{year}{2018}).

\bibitem{Kittlaus2018}
\bibinfo{author}{Kittlaus, E.~A.}, \bibinfo{author}{Otterstrom, N.~T.},
  \bibinfo{author}{Kharel, P.}, \bibinfo{author}{Gertler, S.} \&
  \bibinfo{author}{Rakich, P.~T.}
\newblock \bibinfo{title}{Non-reciprocal interband {Brillouin} modulation}.
\newblock \emph{\bibinfo{journal}{Nat. Photonics}}
  \textbf{\bibinfo{volume}{12}}, \bibinfo{pages}{613--619}
  (\bibinfo{year}{2018}).

\bibitem{Peterson:18}
\bibinfo{author}{Peterson, C.~W.}, \bibinfo{author}{Kim, S.},
  \bibinfo{author}{Bernhard, J.~T.} \& \bibinfo{author}{Bahl, G.}
\newblock \bibinfo{title}{Synthetic phonons enable nonreciprocal coupling to
  arbitrary resonator networks}.
\newblock \emph{\bibinfo{journal}{Science Advances}}
  \textbf{\bibinfo{volume}{4}}, \bibinfo{pages}{eaat0232}
  (\bibinfo{year}{2018}).

\bibitem{Shi:2018aa}
\bibinfo{author}{Shi, Y.}, \bibinfo{author}{Lin, Q.}, \bibinfo{author}{Minkov,
  M.} \& \bibinfo{author}{Fan, S.}
\newblock \bibinfo{title}{Nonreciprocal optical dissipation based on
  direction-dependent rabi splitting}.
\newblock \emph{\bibinfo{journal}{IEEE Journal of Selected Topics in Quantum
  Electronics}} \textbf{\bibinfo{volume}{24}}, \bibinfo{pages}{1--7}
  (\bibinfo{year}{2018}).

\bibitem{Sohn:2019aa}
\bibinfo{author}{Sohn, D.~B.} \& \bibinfo{author}{Bahl, G.}
\newblock \bibinfo{title}{Direction reconfigurable nonreciprocal acousto-optic
  modulator on chip}.
\newblock \emph{\bibinfo{journal}{APL Photonics}} \textbf{\bibinfo{volume}{4}},
  \bibinfo{pages}{126103} (\bibinfo{year}{2019}).

\bibitem{Tian:2020ti}
\bibinfo{author}{Tian, H.} \emph{et~al.}
\newblock \bibinfo{title}{Hybrid integrated photonics using bulk acoustic
  resonators}.
\newblock \emph{\bibinfo{journal}{Nature Communications}}
  \textbf{\bibinfo{volume}{11}}, \bibinfo{pages}{3073} (\bibinfo{year}{2020}).

\bibitem{sarabalis2020}
\bibinfo{author}{Sarabalis, C.~J.} \emph{et~al.}
\newblock \bibinfo{title}{Acousto-optic modulation of a wavelength-scale
  waveguide}.
\newblock \emph{\bibinfo{journal}{arXiv:2010.12517}}  (\bibinfo{year}{2020}).

\bibitem{Dostart:2021uh}
\bibinfo{author}{Dostart, N.}, \bibinfo{author}{Gevorgyan, H.},
  \bibinfo{author}{Onural, D.} \& \bibinfo{author}{Popovi{\'c}, M.}
\newblock \bibinfo{title}{Optical isolation using microring modulators}.
\newblock \emph{\bibinfo{journal}{Optics Letters}}
  \textbf{\bibinfo{volume}{46}}, \bibinfo{pages}{460--463}
  (\bibinfo{year}{2021}).

\bibitem{Kim:2021te}
\bibinfo{author}{Kim, S.}, \bibinfo{author}{Sohn, D.~B.},
  \bibinfo{author}{Peterson, C.~W.} \& \bibinfo{author}{Bahl, G.}
\newblock \bibinfo{title}{On-chip optical non-reciprocity through a synthetic
  hall effect for photons}.
\newblock \emph{\bibinfo{journal}{APL Photonics}} \textbf{\bibinfo{volume}{6}},
  \bibinfo{pages}{011301} (\bibinfo{year}{2021}).

\bibitem{Kittlaus:2021wq}
\bibinfo{author}{Kittlaus, E.~A.} \emph{et~al.}
\newblock \bibinfo{title}{Electrically driven acousto-optics and broadband
  non-reciprocity in silicon photonics}.
\newblock \emph{\bibinfo{journal}{Nature Photonics}}
  \textbf{\bibinfo{volume}{15}}, \bibinfo{pages}{43--52}
  (\bibinfo{year}{2021}).

\bibitem{Peng:2014ui}
\bibinfo{author}{Peng, B.}, \bibinfo{author}{{\"O}zdemir, {\c S}.~K.},
  \bibinfo{author}{Chen, W.}, \bibinfo{author}{Nori, F.} \&
  \bibinfo{author}{Yang, L.}
\newblock \bibinfo{title}{What is and what is not electromagnetically induced
  transparency in whispering-gallery microcavities}.
\newblock \emph{\bibinfo{journal}{Nature Communications}}
  \textbf{\bibinfo{volume}{5}}, \bibinfo{pages}{5082} (\bibinfo{year}{2014}).

\bibitem{Zhang:2019ww}
\bibinfo{author}{Zhang, M.} \emph{et~al.}
\newblock \bibinfo{title}{Electronically programmable photonic molecule}.
\newblock \emph{\bibinfo{journal}{Nature Photonics}}
  \textbf{\bibinfo{volume}{13}}, \bibinfo{pages}{36--40}
  (\bibinfo{year}{2019}).

\bibitem{Ross:11}
\bibinfo{author}{Bi, L.} \emph{et~al.}
\newblock \bibinfo{title}{On-chip optical isolation in monolithically
  integrated non-reciprocal optical resonators}.
\newblock \emph{\bibinfo{journal}{Nat. Photonics}}
  \textbf{\bibinfo{volume}{5}}, \bibinfo{pages}{758} (\bibinfo{year}{2011}).

\bibitem{Ghosh:12aa}
\bibinfo{author}{Ghosh, S.} \emph{et~al.}
\newblock \bibinfo{title}{{Ce:YIG/Silicon}-on-insulator waveguide optical
  isolator realized by adhesive bonding}.
\newblock \emph{\bibinfo{journal}{Opt. Express}} \textbf{\bibinfo{volume}{20}},
  \bibinfo{pages}{1839--1848} (\bibinfo{year}{2012}).

\bibitem{Huang:17}
\bibinfo{author}{Huang, D.} \emph{et~al.}
\newblock \bibinfo{title}{Dynamically reconfigurable integrated optical
  circulators}.
\newblock \emph{\bibinfo{journal}{Optica}} \textbf{\bibinfo{volume}{4}},
  \bibinfo{pages}{23--30} (\bibinfo{year}{2017}).

\bibitem{Zhang:2017wq}
\bibinfo{author}{Zhang, C.}, \bibinfo{author}{Dulal, P.},
  \bibinfo{author}{Stadler, B. J.~H.} \& \bibinfo{author}{Hutchings, D.~C.}
\newblock \bibinfo{title}{Monolithically-integrated {TE}-mode {1D}
  silicon-on-insulator isolators using seedlayer-free garnet}.
\newblock \emph{\bibinfo{journal}{Scientific Reports}}
  \textbf{\bibinfo{volume}{7}}, \bibinfo{pages}{5820} (\bibinfo{year}{2017}).

\bibitem{Du:2018wy}
\bibinfo{author}{Du, Q.} \emph{et~al.}
\newblock \bibinfo{title}{Monolithic on-chip magneto-optical isolator with 3
  {dB} insertion loss and 40 db isolation ratio}.
\newblock \emph{\bibinfo{journal}{ACS Photonics}} \textbf{\bibinfo{volume}{5}},
  \bibinfo{pages}{5010--5016} (\bibinfo{year}{2018}).

\bibitem{Zhang:19}
\bibinfo{author}{Zhang, Y.} \emph{et~al.}
\newblock \bibinfo{title}{Monolithic integration of broadband optical isolators
  for polarization-diverse silicon photonics}.
\newblock \emph{\bibinfo{journal}{Optica}} \textbf{\bibinfo{volume}{6}},
  \bibinfo{pages}{473--478} (\bibinfo{year}{2019}).

\bibitem{Yan:20}
\bibinfo{author}{Yan, W.} \emph{et~al.}
\newblock \bibinfo{title}{Waveguide-integrated high-performance magneto-optical
  isolators and circulators on silicon nitride platforms}.
\newblock \emph{\bibinfo{journal}{Optica}} \textbf{\bibinfo{volume}{7}},
  \bibinfo{pages}{1555--1562} (\bibinfo{year}{2020}).

\bibitem{Maayani:2018vc}
\bibinfo{author}{Maayani, S.} \emph{et~al.}
\newblock \bibinfo{title}{Flying couplers above spinning resonators generate
  irreversible refraction}.
\newblock \emph{\bibinfo{journal}{Nature}} \textbf{\bibinfo{volume}{558}},
  \bibinfo{pages}{569--572} (\bibinfo{year}{2018}).

\bibitem{Scheucher:16}
\bibinfo{author}{Scheucher, M.}, \bibinfo{author}{Hilico, A.},
  \bibinfo{author}{Will, E.}, \bibinfo{author}{Volz, J.} \&
  \bibinfo{author}{Rauschenbeutel, A.}
\newblock \bibinfo{title}{Quantum optical circulator controlled by a single
  chirally coupled atom}.
\newblock \emph{\bibinfo{journal}{Science}} \textbf{\bibinfo{volume}{354}},
  \bibinfo{pages}{1577} (\bibinfo{year}{2016}).

\bibitem{Spencer:2018ta}
\bibinfo{author}{Spencer, D.~T.} \emph{et~al.}
\newblock \bibinfo{title}{An optical-frequency synthesizer using integrated
  photonics}.
\newblock \emph{\bibinfo{journal}{Nature}} \textbf{\bibinfo{volume}{557}},
  \bibinfo{pages}{81--85} (\bibinfo{year}{2018}).

\bibitem{Lucas:2020tz}
\bibinfo{author}{Lucas, E.} \emph{et~al.}
\newblock \bibinfo{title}{Ultralow-noise photonic microwave synthesis using a
  soliton microcomb-based transfer oscillator}.
\newblock \emph{\bibinfo{journal}{Nature Communications}}
  \textbf{\bibinfo{volume}{11}}, \bibinfo{pages}{374} (\bibinfo{year}{2020}).

\bibitem{Poulton:2017tn}
\bibinfo{author}{Poulton, C.~V.} \emph{et~al.}
\newblock \bibinfo{title}{Coherent solid-state lidar with silicon photonic
  optical phased arrays}.
\newblock \emph{\bibinfo{journal}{Optics Letters}}
  \textbf{\bibinfo{volume}{42}}, \bibinfo{pages}{4091--4094}
  (\bibinfo{year}{2017}).

\bibitem{DelHaye:2007tv}
\bibinfo{author}{Del'Haye, P.} \emph{et~al.}
\newblock \bibinfo{title}{Optical frequency comb generation from a monolithic
  microresonator}.
\newblock \emph{\bibinfo{journal}{Nature}} \textbf{\bibinfo{volume}{450}},
  \bibinfo{pages}{1214--1217} (\bibinfo{year}{2007}).

\bibitem{Hummon:2018wp}
\bibinfo{author}{Hummon, M.~T.} \emph{et~al.}
\newblock \bibinfo{title}{Photonic chip for laser stabilization to an atomic
  vapor with 10$^{-11}$ instability}.
\newblock \emph{\bibinfo{journal}{Optica}} \textbf{\bibinfo{volume}{5}},
  \bibinfo{pages}{443--449} (\bibinfo{year}{2018}).

\bibitem{Knappe:04}
\bibinfo{author}{Knappe, S.} \emph{et~al.}
\newblock \bibinfo{title}{A microfabricated atomic clock}.
\newblock \emph{\bibinfo{journal}{Appl. Phys. Lett.}}
  \textbf{\bibinfo{volume}{85}}, \bibinfo{pages}{1460--1462}
  (\bibinfo{year}{2004}).

\bibitem{Newman:2019wo}
\bibinfo{author}{Newman, Z.~L.} \emph{et~al.}
\newblock \bibinfo{title}{Architecture for the photonic integration of an
  optical atomic clock}.
\newblock \emph{\bibinfo{journal}{Optica}} \textbf{\bibinfo{volume}{6}},
  \bibinfo{pages}{680--685} (\bibinfo{year}{2019}).

\bibitem{Gong:13}
\bibinfo{author}{{Gong}, S.} \& \bibinfo{author}{{Piazza}, G.}
\newblock \bibinfo{title}{Design and analysis of lithium--niobate-based high
  electromechanical coupling {RF-MEMS} resonators for wideband filtering}.
\newblock \emph{\bibinfo{journal}{IEEE Transactions on Microwave Theory and
  Techniques}} \textbf{\bibinfo{volume}{61}}, \bibinfo{pages}{403--414}
  (\bibinfo{year}{2013}).

\bibitem{Riedel:2010te}
\bibinfo{author}{Riedel, M.~F.} \emph{et~al.}
\newblock \bibinfo{title}{Atom-chip-based generation of entanglement for
  quantum metrology}.
\newblock \emph{\bibinfo{journal}{Nature}} \textbf{\bibinfo{volume}{464}},
  \bibinfo{pages}{1170--1173} (\bibinfo{year}{2010}).

\bibitem{Blanshan:05}
\bibinfo{author}{Blanshan, E.}, \bibinfo{author}{Rochester, S.~M.},
  \bibinfo{author}{Donley, E.~A.} \& \bibinfo{author}{Kitching, J.}
\newblock \bibinfo{title}{Light shifts in a pulsed cold-atom
  coherent-population-trapping clock}.
\newblock \emph{\bibinfo{journal}{Phys. Rev. A}} \textbf{\bibinfo{volume}{91}},
  \bibinfo{pages}{041401} (\bibinfo{year}{2015}).

\end{thebibliography}


\begin{thebibliography}{10}
\expandafter\ifx\csname url\endcsname\relax
  \def\url#1{\texttt{#1}}\fi
\expandafter\ifx\csname urlprefix\endcsname\relax\def\urlprefix{URL }\fi
\providecommand{\bibinfo}[2]{#2}
\providecommand{\eprint}[2][]{\url{#2}}

\bibitem{Sohn18}
\bibinfo{author}{Sohn, D.~B.}, \bibinfo{author}{Kim, S.} \&
  \bibinfo{author}{Bahl, G.}
\newblock \bibinfo{title}{Time-reversal symmetry breaking with acoustic pumping
  of nanophotonic circuits}.
\newblock \emph{\bibinfo{journal}{Nat. Photonics}}
  \textbf{\bibinfo{volume}{12}}, \bibinfo{pages}{91--97}
  (\bibinfo{year}{2018}).

\bibitem{Cai:00}
\bibinfo{author}{Cai, M.}, \bibinfo{author}{Painter, O.} \&
  \bibinfo{author}{Vahala, K.~J.}
\newblock \bibinfo{title}{Observation of critical coupling in a fiber taper to
  a silica-microsphere whispering-gallery mode system}.
\newblock \emph{\bibinfo{journal}{Phys. Rev. Lett.}}
  \textbf{\bibinfo{volume}{85}}, \bibinfo{pages}{74--77}
  (\bibinfo{year}{2000}).

\bibitem{Kuznetsova:01}
\bibinfo{author}{{Kuznetsova}, I.~E.}, \bibinfo{author}{{Zaitsev}, B.~D.},
  \bibinfo{author}{{Joshi}, S.~G.} \& \bibinfo{author}{{Borodina}, I.~A.}
\newblock \bibinfo{title}{Investigation of acoustic waves in thin plates of
  lithium niobate and lithium tantalate}.
\newblock \emph{\bibinfo{journal}{IEEE Transactions on Ultrasonics,
  Ferroelectrics, and Frequency Control}} \textbf{\bibinfo{volume}{48}},
  \bibinfo{pages}{322--328} (\bibinfo{year}{2001}).

\bibitem{Sohn:2019aa}
\bibinfo{author}{Sohn, D.~B.} \& \bibinfo{author}{Bahl, G.}
\newblock \bibinfo{title}{Direction reconfigurable nonreciprocal acousto-optic
  modulator on chip}.
\newblock \emph{\bibinfo{journal}{APL Photonics}} \textbf{\bibinfo{volume}{4}},
  \bibinfo{pages}{126103} (\bibinfo{year}{2019}).

\bibitem{Kittlaus:2021wq}
\bibinfo{author}{Kittlaus, E.~A.} \emph{et~al.}
\newblock \bibinfo{title}{Electrically driven acousto-optics and broadband
  non-reciprocity in silicon photonics}.
\newblock \emph{\bibinfo{journal}{Nature Photonics}}
  \textbf{\bibinfo{volume}{15}}, \bibinfo{pages}{43--52}
  (\bibinfo{year}{2021}).

\bibitem{Kittlaus2018}
\bibinfo{author}{Kittlaus, E.~A.}, \bibinfo{author}{Otterstrom, N.~T.},
  \bibinfo{author}{Kharel, P.}, \bibinfo{author}{Gertler, S.} \&
  \bibinfo{author}{Rakich, P.~T.}
\newblock \bibinfo{title}{Non-reciprocal interband {Brillouin} modulation}.
\newblock \emph{\bibinfo{journal}{Nat. Photonics}}
  \textbf{\bibinfo{volume}{12}}, \bibinfo{pages}{613--619}
  (\bibinfo{year}{2018}).

\bibitem{sarabalis2020}
\bibinfo{author}{Sarabalis, C.~J.} \emph{et~al.}
\newblock \bibinfo{title}{Acousto-optic modulation of a wavelength-scale
  waveguide}.
\newblock \emph{\bibinfo{journal}{arXiv:2010.12517}}  (\bibinfo{year}{2020}).

\bibitem{Liu:19}
\bibinfo{author}{Liu, Q.}, \bibinfo{author}{Li, H.} \& \bibinfo{author}{Li, M.}
\newblock \bibinfo{title}{Electromechanical {Brillouin} scattering in
  integrated optomechanical waveguides}.
\newblock \emph{\bibinfo{journal}{Optica}} \textbf{\bibinfo{volume}{6}},
  \bibinfo{pages}{778--785} (\bibinfo{year}{2019}).

\bibitem{Kim2016}
\bibinfo{author}{Kim, J.}, \bibinfo{author}{Kim, S.} \& \bibinfo{author}{Bahl,
  G.}
\newblock \bibinfo{title}{Complete linear optical isolation at the microscale
  with ultralow loss}.
\newblock \emph{\bibinfo{journal}{Sci. Rep.}} \textbf{\bibinfo{volume}{7}},
  \bibinfo{pages}{1647} (\bibinfo{year}{2017}).

\bibitem{Kim:2021te}
\bibinfo{author}{Kim, S.}, \bibinfo{author}{Sohn, D.~B.},
  \bibinfo{author}{Peterson, C.~W.} \& \bibinfo{author}{Bahl, G.}
\newblock \bibinfo{title}{On-chip optical non-reciprocity through a synthetic
  hall effect for photons}.
\newblock \emph{\bibinfo{journal}{APL Photonics}} \textbf{\bibinfo{volume}{6}},
  \bibinfo{pages}{011301} (\bibinfo{year}{2021}).

\bibitem{Dostart:2021uh}
\bibinfo{author}{Dostart, N.}, \bibinfo{author}{Gevorgyan, H.},
  \bibinfo{author}{Onural, D.} \& \bibinfo{author}{Popovi{\'c}, M.}
\newblock \bibinfo{title}{Optical isolation using microring modulators}.
\newblock \emph{\bibinfo{journal}{Optics Letters}}
  \textbf{\bibinfo{volume}{46}}, \bibinfo{pages}{460--463}
  (\bibinfo{year}{2021}).

\bibitem{Doerr:14}
\bibinfo{author}{Doerr, C.~R.}, \bibinfo{author}{Chen, L.} \&
  \bibinfo{author}{Vermeulen, D.}
\newblock \bibinfo{title}{Silicon photonics broadband modulation-based
  isolator}.
\newblock \emph{\bibinfo{journal}{Opt. Express}} \textbf{\bibinfo{volume}{22}},
  \bibinfo{pages}{4493--4498} (\bibinfo{year}{2014}).

\bibitem{Tzuang2014}
\bibinfo{author}{Tzuang, L.~D.}, \bibinfo{author}{Fang, K.},
  \bibinfo{author}{Nussenzveig, P.}, \bibinfo{author}{Fan, S.} \&
  \bibinfo{author}{Lipson, M.}
\newblock \bibinfo{title}{Non-reciprocal phase shift induced by an effective
  magnetic flux for light}.
\newblock \emph{\bibinfo{journal}{Nat. Photonics}}
  \textbf{\bibinfo{volume}{8}}, \bibinfo{pages}{701} (\bibinfo{year}{2014}).

\bibitem{lira2012}
\bibinfo{author}{Lira, H.}, \bibinfo{author}{Yu, Z.}, \bibinfo{author}{Fan, S.}
  \& \bibinfo{author}{Lipson, M.}
\newblock \bibinfo{title}{Electrically driven nonreciprocity induced by
  interband photonic transition on a silicon chip}.
\newblock \emph{\bibinfo{journal}{Phys. Rev. Lett.}}
  \textbf{\bibinfo{volume}{109}}, \bibinfo{pages}{033901}
  (\bibinfo{year}{2012}).

\bibitem{Doerr:11}
\bibinfo{author}{Doerr, C.~R.}, \bibinfo{author}{Dupuis, N.} \&
  \bibinfo{author}{Zhang, L.}
\newblock \bibinfo{title}{Optical isolator using two tandem phase modulators}.
\newblock \emph{\bibinfo{journal}{Opt. Lett.}} \textbf{\bibinfo{volume}{36}},
  \bibinfo{pages}{4293--4295} (\bibinfo{year}{2011}).

\bibitem{Yan:20}
\bibinfo{author}{Yan, W.} \emph{et~al.}
\newblock \bibinfo{title}{Waveguide-integrated high-performance magneto-optical
  isolators and circulators on silicon nitride platforms}.
\newblock \emph{\bibinfo{journal}{Optica}} \textbf{\bibinfo{volume}{7}},
  \bibinfo{pages}{1555--1562} (\bibinfo{year}{2020}).

\bibitem{Pintus:19}
\bibinfo{author}{{Pintus}, P.} \emph{et~al.}
\newblock \bibinfo{title}{Broadband te optical isolators and circulators in
  silicon photonics through ce:yig bonding}.
\newblock \emph{\bibinfo{journal}{Journal of Lightwave Technology}}
  \textbf{\bibinfo{volume}{37}}, \bibinfo{pages}{1463--1473}
  (\bibinfo{year}{2019}).

\bibitem{Zhang:19}
\bibinfo{author}{Zhang, Y.} \emph{et~al.}
\newblock \bibinfo{title}{Monolithic integration of broadband optical isolators
  for polarization-diverse silicon photonics}.
\newblock \emph{\bibinfo{journal}{Optica}} \textbf{\bibinfo{volume}{6}},
  \bibinfo{pages}{473--478} (\bibinfo{year}{2019}).

\bibitem{Du:2018wy}
\bibinfo{author}{Du, Q.} \emph{et~al.}
\newblock \bibinfo{title}{Monolithic on-chip magneto-optical isolator with 3
  {dB} insertion loss and 40 db isolation ratio}.
\newblock \emph{\bibinfo{journal}{ACS Photonics}} \textbf{\bibinfo{volume}{5}},
  \bibinfo{pages}{5010--5016} (\bibinfo{year}{2018}).

\bibitem{Ross:11}
\bibinfo{author}{Bi, L.} \emph{et~al.}
\newblock \bibinfo{title}{On-chip optical isolation in monolithically
  integrated non-reciprocal optical resonators}.
\newblock \emph{\bibinfo{journal}{Nat. Photonics}}
  \textbf{\bibinfo{volume}{5}}, \bibinfo{pages}{758} (\bibinfo{year}{2011}).

\bibitem{Huang:17}
\bibinfo{author}{Huang, D.} \emph{et~al.}
\newblock \bibinfo{title}{Dynamically reconfigurable integrated optical
  circulators}.
\newblock \emph{\bibinfo{journal}{Optica}} \textbf{\bibinfo{volume}{4}},
  \bibinfo{pages}{23--30} (\bibinfo{year}{2017}).

\bibitem{Ghosh:12aa}
\bibinfo{author}{Ghosh, S.} \emph{et~al.}
\newblock \bibinfo{title}{{Ce:YIG/Silicon}-on-insulator waveguide optical
  isolator realized by adhesive bonding}.
\newblock \emph{\bibinfo{journal}{Opt. Express}} \textbf{\bibinfo{volume}{20}},
  \bibinfo{pages}{1839--1848} (\bibinfo{year}{2012}).

\end{thebibliography}
\end{bibunit}
\section*{Acknowledgments}

This work was sponsored by the Defense Advanced Research Projects Agency (DARPA) grant FA8650-19-2-7924, the National Science Foundation EFRI grant EFMA-1641084, and the Air Force Office of Scientific Research (AFOSR) grant FA9550-19-1-0256. GB would additionally like to acknowledge support from the Office of Naval Research (ONR) Director for Research Early Career grant N00014-17-1-2209, and the Presidential Early Career Award for Scientists and Engineers. DBS would also like to acknowledge support from a US National Science Foundation Graduate Research Fellowship. The US Government is authorized to reproduce and distribute reprints for Governmental purposes notwithstanding any copyright notation thereon. The views and conclusions contained herein are those of the authors and should not be interpreted as necessarily representing the official policies or endorsements, either expressed or implied, of DARPA or the U.S. Government.

\section*{Data availability}

The data that support the findings of this study are available from the corresponding author upon reasonable request.

\FloatBarrier

\newpage

\renewcommand*{\citenumfont}[1]{S#1}
\renewcommand*{\bibnumfmt}[1]{[S#1]}
\newcommand{\beginsupplement}{%
        \setcounter{table}{0}
        \renewcommand{\thetable}{S\arabic{table}}%
        \setcounter{figure}{0}
        \renewcommand{\thefigure}{S\arabic{figure}}%
        \setcounter{equation}{0}
        \renewcommand{\theequation}{S\arabic{equation}}
        \setcounter{section}{0}
        \renewcommand{\thesection}{S\arabic{section}}%
}

\beginsupplement
\begin{bibunit}

\begin{center}

\Large{\textbf{Supplementary Information: \\  Electrically driven linear optical isolation through phonon mediated Autler-Townes splitting}} \\
\vspace{12pt}
\vspace{12pt}
\large{{Donggyu B. Sohn}$^{\dag}$,
{Oğulcan E. Örsel}$^{\dag}$,
and {Gaurav Bahl}} \\
\vspace{12pt}
    \footnotesize{Department of Mechanical Science and Engineering,} \\
    \footnotesize{University of Illinois at Urbana–Champaign, Urbana, IL 61801 USA,} \\
    \footnotesize{$\dag$ These authors contributed equally}\\
\end{center}

\vspace{24pt}


\section{Analytical description of phonon mediated ATS}
\label{sec:ATSTheory}

\begin{figure}[tb]
    \begin{adjustwidth}{-1in}{-1in}
    \centering
    \includegraphics[width=0.4\textwidth]{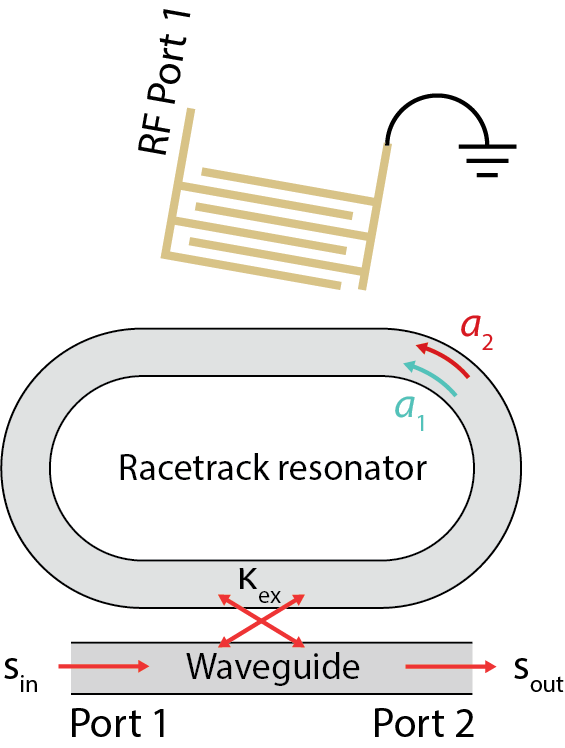}
    \caption{Illustration of variables of interest.
        }
    \label{sup:variables}
    \end{adjustwidth}
\end{figure}

Here we present a mathematical discussion of the phonon mediated ATS process by invoking the analysis previously performed in \cite{Sohn18}.
The coupled equations of motion for the optical fields in the WGR and waveguide (see Fig.~\ref{sup:variables}) can be expressed in matrix form:
\begin{align}
    \begin{pmatrix}
        \dot{a}_1\\
        \dot{a}_2
    \end{pmatrix}
        =
    \begin{pmatrix}
        -i\omega_1-\cfrac{\kappa_1}{2}   &  i \left( \cfrac{G_{ph}}{2} \right) e^{i\Omega t}\\
        i \left( \cfrac{G_{ph}}{2} \right) e^{-i\Omega t} & -i\omega_2-\cfrac{\kappa_2}{2}
    \end{pmatrix}
    \begin{pmatrix}
        a_1\\
        a_2
    \end{pmatrix}
        +
    \begin{pmatrix}
        \sqrt{\kappa_{ex1}} \\
        \sqrt{\kappa_{ex2}} 
    \end{pmatrix}
    s_{in}e^{-i\omega_lt}
\end{align}
where $a_1$ ($a_2$) is the intracavity field for the TE$_{00}$ (TE$_{10}$) mode, $\kappa_1$ ($\kappa_2$) is the loaded decay rate, $\kappa_{ex1}$ ($\kappa_{ex2}$) is the external coupling rate, $\omega_1$ ($\omega_2$) is the resonant frequency, $G_{ph}$ is the acousto-optic coupling rate, $\Omega$ is the RF driving frequency, $\omega_l$ is the input laser frequency, and $s_{in}$ is the input field in the waveguide. 

In our case, the TE$_{00}$ mode is intentionally undercoupled relative to the waveguide in order to minimize the sidebands as discussed below in \S\ref{sec:Sidebands}. Thus, we can neglect the external coupling rate of TE$_{00}$ mode (set $\kappa_{ex1} = 0$). 
We can define an additional subscript for the field terms $a_{2,0}$ to indicate the original carrier frequency, $a_{2,+1}$ to indicate the first order upper sideband, and $a_{2,-1}$ to indicate the first order lower sideband respectively.
Presently, we are interested in the optical output at the input (carrier) frequency $s_{out,0}$  (ignoring the sidebands $s_{out,\pm 1}$) from the waveguide using the expression: 
\begin{align}
    \label{eqn:outputeqn}
    s_{out,0} = s_{in} - \sqrt{\kappa_{ex2}} \, a_{2,0}
\end{align}
where the TE$_{10}$ intracavity field at the input frequency ($a_{2,0}$) can be expressed as:
\begin{align}
    \label{eqn:a2_phasematched}
    a_{2,0} = 
        \frac
            {\sqrt{\kappa_{ex2}} \, \left( \cfrac{\kappa_1}{2} - i(\Delta - \delta) \right)}
            { \left( \cfrac{\kappa_1}{2} - i(\Delta - \delta) \right) \left( \cfrac{\kappa_1}{2} - i\Delta \right) + \cfrac{G^2_{ph}}{4} } 
        \, s_{in}
\end{align}
where $\Delta = \omega_l - \omega_2$ is the optical detuning from the TE$_{10}$ mode and $\delta = \omega_1 - \omega_2 +\Omega$ is the frequency mismatch between the optical mode separation and the applied phonon frequency. Additional details on the upper/lower sidebands can be found in \cite{Sohn18}, but as we discuss below in \S\ref{sec:Sidebands}, they play a relatively minor role in this system.

For simplicity, we now assume that the loss rates of the optical modes are the same ($\kappa_1 = \kappa_2 = \kappa$) and their frequency difference is perfectly matched to the applied phonons ($\delta = 0$). We can then simplify the TE$_{10}$ intracavity field using two diagonalized intracavity fields as follows: 
\begin{align}
    \label{eqn:a2_split}
    a_{2,0} = 
        \left[  \frac{\sqrt{\kappa_{ex2}}/2}{\nicefrac{\kappa}{2} - i \left(\Delta+ \nicefrac{G_{ph}}{2} \right)} 
                + \frac{\sqrt{\kappa_{ex2}}/2}{\nicefrac{\kappa}{2} - i \left(\Delta- \nicefrac{G_{ph}}{2} \right)}
        \right] s_{in}
\end{align}
Here we can easily see that the cavity susceptibility (the expression in the square brackets) is now modified from a single Lorentzian response, and instead exhibits two Lorentzian shaped responses corresponding to the dressed states that are split by the Rabi frequency $G_{ph}$.
If the separation is large enough ($G_{ph} \gg \kappa$), there is essentially zero optical density of states at the original TE$_{10}$ resonant frequency ($\omega_2$). In this situation, $a_{2,0} \rightarrow 0$ and $s_{out,0} \rightarrow s_{in}$, i.e. light in the waveguide transmits with unity gain. 
This conclusion about transparency remains true irrespective of the value of the extrinsic coupling coefficient $\kappa_{ex2}$.
As discussed in the main text (Fig.~\ref{fig:1}), this ATS process is momentum sensitive and only exists for one circulation direction around the resonator.

\section{Independent control of isolation contrast via critical coupling}
\label{sup:critical}

In the non-phase matched direction, where the optomechanical coupling does not exist (i.e. $G_{ph} = 0$), the optical equations of motion can be rewritten:
\begin{align}
    \begin{pmatrix}
        \dot{a}_1\\
        \dot{a}_2
    \end{pmatrix}
        =
    \begin{pmatrix}
        -i\omega_1-\cfrac{\kappa_1}{2}   &  0\\
        0 & -i\omega_2-\cfrac{\kappa_2}{2}
    \end{pmatrix}
    \begin{pmatrix}
        a_1\\
        a_2
    \end{pmatrix}
        +
    \begin{pmatrix}
        \sqrt{\kappa_{ex1}} \\
        \sqrt{\kappa_{ex2}} 
    \end{pmatrix}
    s_{in}e^{-i\omega_lt}
\end{align}
Again setting $\kappa_{ex1} = 0$, the output field $s_{out,0}$ can be expressed as: 
\begin{align}
    s_{out,0} = s_{in} - \sqrt{\kappa_{ex2}} \, a_{2,0} = s_{in}\left(1-\cfrac{\kappa_{ex2}}{ \cfrac{\kappa_2}{2} - i\Delta}\right)
\end{align}
From this equation, we can find that the transmission vanishes when $\kappa_2 / 2 = \kappa_{ex2}$ at the original resonance frequency ($\Delta = 0$). Since the total loss rate for an optical mode is the sum of intrinsic loss rate and the external loss rate ($\kappa_2 = \kappa_{int2} + \kappa_{ex2}$), we find that critical coupling occurs when the intrinsic and external loss rates are the same $\kappa_{int2} = \kappa_{ex2}$~\cite{Cai:00}.
As discussed in \S\ref{sec:ATSTheory}, the transparency is independent (to first order) of the value of $\kappa_{ex2}$. As a result, barring practical constraints, we can independently set the isolation contrast to as large a value as desired by optimizing the system to reach the critical coupling point for the TE$_{10}$ mode.

\section{Suppression of modulation sidebands}
\label{sec:Sidebands}

A general concern with non-reciprocal systems based on spatio-temporal modulation is the generation of sidebands. Even though these sidebands are typically separated from the carrier by multiples of the modulation frequency, e.g. several GHz, they are generally undesirable as they can require filtering and are essentially sources of loss out from the optical modes of interest. 
In the system we propose, sideband generation is naturally suppressed and only linear optical isolation is produced. To understand this, we recall the TE$_{00}$ sideband amplitude, which is non-zero due to light coupling in from the TE$_{10}$ mode~\cite{Sohn18}. Here, only Stokes sideband presents due to the phase matching condition.
\begin{align}
    s_{out,-1} &= \sqrt{\kappa_{ex2}} \, a_{1,-1}  \notag \\
    &= \frac{-i\cfrac{G_{ph}}{2}\sqrt{\kappa_{ex1}\kappa_{ex2}}}{\left(\cfrac{\kappa_1}{2} - i(\Delta - \delta)\right)\left(\cfrac{\kappa_2}{2} - i\Delta\right) + \cfrac{G_{ph^2}}{4} } s_{in}
\end{align}

\begin{figure}[tb]
    \begin{adjustwidth}{-1in}{-1in}
    \centering
    \includegraphics[width=1.1\textwidth]{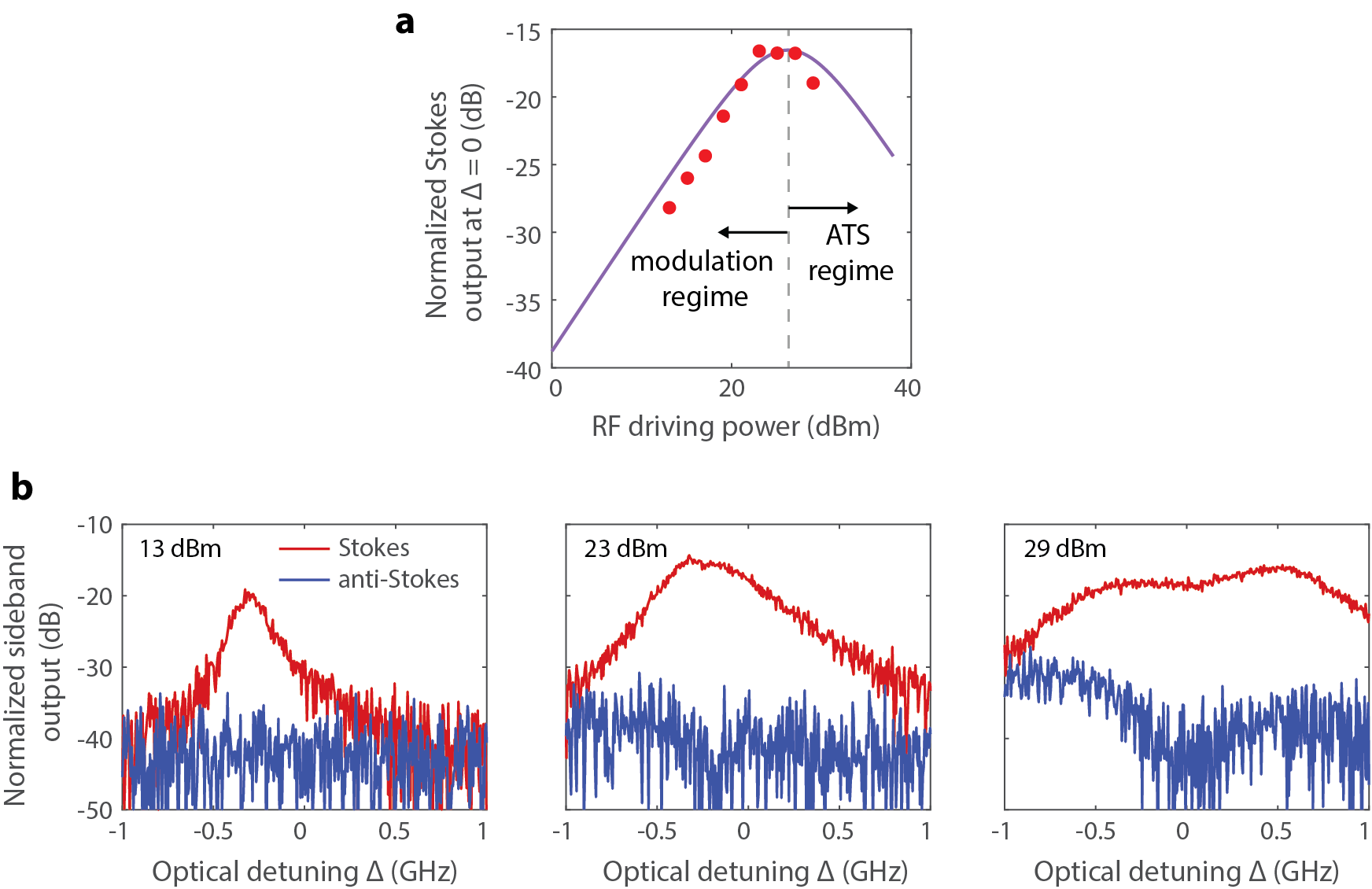}
    \caption{
        \textbf{(a)} Normalized Stokes sideband power (Stokes power divided by Carrier power) for the 1550 nm device shown in Fig.~\ref{fig:3}a, at the zero detuning point $\Delta = 0$. As the optomechanical coupling increases ($G_{ph} \propto \sqrt{P_{RF}}$), we find that the Stokes sideband initially rises due to increase acousto-optic coupling. However, once the ATS mode split is more fully resolved, the sideband power decreases due to the reduced optical density of states near zero detuning. The solid line is a model prediction using the experimentally measured device parameters.
        \textbf{(b)} Full traces of the Stokes and anti-Stokes sidebands as a function of the optical detuning. The anti-Stokes sideband is always very small due to the absence of phase matching.
    }
    \label{sup:sideband}
    \end{adjustwidth}
\end{figure}

From this equation, we can observe that the sideband generation can be suppressed with two different methods. The first option is to reduce the external coupling rate for the TE$_{00}$ optical mode ($\kappa_{ex1} \approx 0$). In this case, the scattered light no longer has an exit path via the waveguide and is just dissipated in the resonator. This can be achieved by carefully engineering the waveguide and resonator, and is naturally easy since the TE$_{10}$ mode has a larger evanescent field than the TE$_{00}$ mode. 
The second option is to increase the phonon-induced optomechanical coupling ($G_{ph} \gg \kappa$). In this case, the separated optical modes are fully resolved and no light couples into the TE$_{10}$ mode at its original resonance (see Eqn.~\ref{eqn:a2_split}). Thus, there is no light to be scattered. 

In the experimental results shown in Fig.~\ref{sup:sideband}, we demonstrate the effect of increasing $G_{ph}$ by showing the normalized sideband amplitude for different RF drive levels. 
As $G_{ph}$ (RF input power) increases, the amplitude of the sideband first increases due to the higher acousto-optic coupling rate as in \cite{Sohn18}. However, as the ATS mode split appears, the density of states vanishes at the original resonant frequency and the sideband is suppressed. This suppression effect gets stronger as $G_{ph}$ is increased further.

\FloatBarrier

\section{Critical coupling of the non-reciprocal dressed states}
\label{sec:WingOptimization}

As shown in Fig.~\ref{fig:4} of the main manuscript, some extremely favorable isolation situations can be engineered using the dressed states TE$^\pm_{10}$ of the photonic atom under the ATS system.
These situations arise when the phonon frequency is not precisely matched to the optical mode separation, and additionally the TE$_{10}$ mode is overcoupled.

We can formally explore this by invoking Eqn.~\ref{eqn:outputeqn} and Eqn.~\ref{eqn:a2_phasematched} to write the waveguide transfer function without any assumptions about the states:
\begin{align}
    \label{eqn:DressedTF}
    \frac{s_{out, 0}}{s_{in}} = 1 -  
        \frac
            {\kappa_{ex2} \, \left( \cfrac{\kappa_1}{2} - i(\Delta - \delta) \right)}
            { \left( \cfrac{\kappa_1}{2} - i(\Delta - \delta) \right) \left( \cfrac{\kappa_2}{2} - i\Delta \right) + \cfrac{G^2_{ph}}{4} } 
        \,
\end{align}
where $\delta = \omega_1 - \omega_2 +\Omega$ is the frequency mismatch between the optical mode separation and the applied phonon frequency. The above transfer function is a consequence of the interaction with both dressed states, and our goal is to find solutions where this transfer function dips to zero.
    
\begin{figure}[htb]
    \begin{adjustwidth}{-1in}{-1in}
    \centering
    \includegraphics[width=0.8\textwidth]{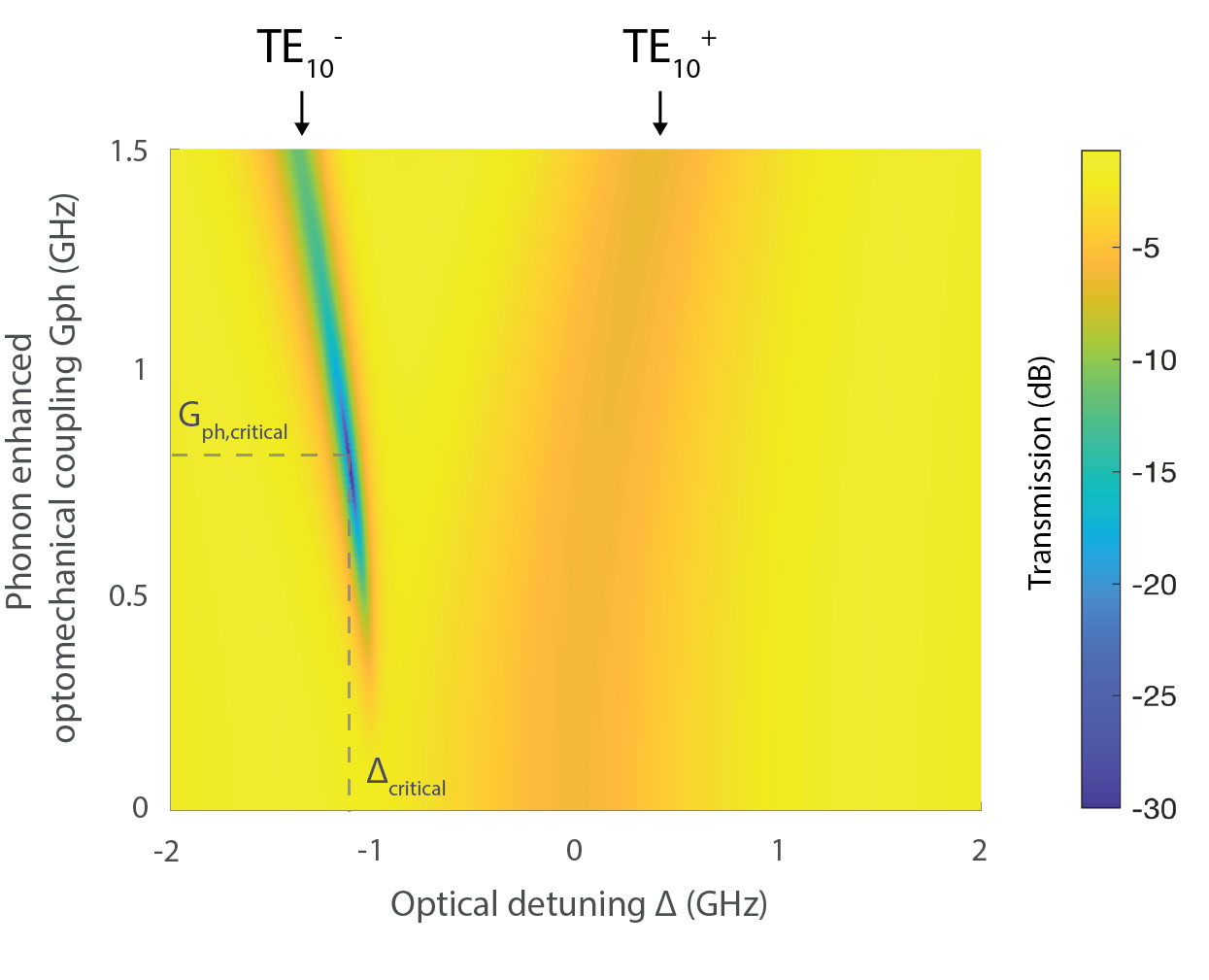}
    \caption{
        Simulation of the transmission $s_{out}/s_{in}$ through the waveguide for detuning $\Delta$ relative to the original TE$_{10}$ mode. The mismatch is set to $\delta = -1$ GHz. We find that a sweet spot exists in the selection of $G_{ph}$ to critically couple the lower dressed state and thereby obtain giant isolation contrast. We additionally note an ac Stark shift phenomenon due to the large mismatch $\delta$. This shift is additionally beneficial for moving the high contrast away from the original TE$_{10}$ mode at $\Delta = 0$, as is seen in the main manuscript Fig.~\ref{fig:4} and in Fig.~\ref{sup:asym780}.
    }
    \label{sup:asymGphsweep}
    \end{adjustwidth}
\end{figure}

To obtain a result, we solve $\nicefrac{s_{out}}{s_{in}} = 0$ by setting the real and imaginary parts of Eqn.~\ref{eqn:DressedTF} independently to zero. The following simultaneous result is obtained:
\begin{align}
    \Delta_{critical} &=     \left(\frac{\kappa_{ex2}-\cfrac{\kappa_2}{2}}{\kappa_{ex2}-\cfrac{\kappa_2}{2}-\cfrac{\kappa_1}{2}} \right) \delta  \label{deltacritical}\\
    G_{ph,critical}^2 &= 4\left(\kappa_{ex2}-\frac{\kappa_2}{2}\right)\left(\frac{\kappa_1}{2}\right)\left(1+\frac{\delta^2}{\left(\kappa_{ex2}-\cfrac{\kappa_2}{2}-\cfrac{\kappa_1}{2}\right)^2}\right) \label{gphcritical}
\end{align}
The above equations indicate that the transmission will dip to zero at the offset frequency $\Delta_{critical}$, which is a fixed characteristic of the fabricated device, when the Rabi frequency hits $G_{ph,critical}$, which is a free parameter that we can control with the RF drive power. 
We also note that $\Delta_{critical}$ simply approaches $\delta$, i.e. the location of the lower dressed state TE$^-_{10}$, if the original TE$_{10}$ mode is highly overcoupled ($\kappa_{ex2}$ dominates) or when the secondary optical mode has very low loss rate ($\kappa_1 \ll \kappa_2$).  
An example simulation showing this criticality condition on the lower dressed state is shown in Fig.~\ref{sup:asymGphsweep}.

More importantly, from Eqn.~\ref{gphcritical}, we find that a real solution for $G_{ph,critical}$ is only possible when $\kappa_{ex2}>\kappa_2/2$, i.e. when the TE$_{10}$ mode is overcoupled.
We also find that the required $G_{ph,critical}$ increases as the frequency mismatch $\delta$ increases. However, as a tremendous advantage, a larger value of $\delta$ permits a much lower insertion loss with high contrast since this critical coupling point is located further away in frequency space from the original absorption frequency.

\begin{figure}[ht]
    \begin{adjustwidth}{-1in}{-1in}
    \centering
    \includegraphics[width=1\textwidth]{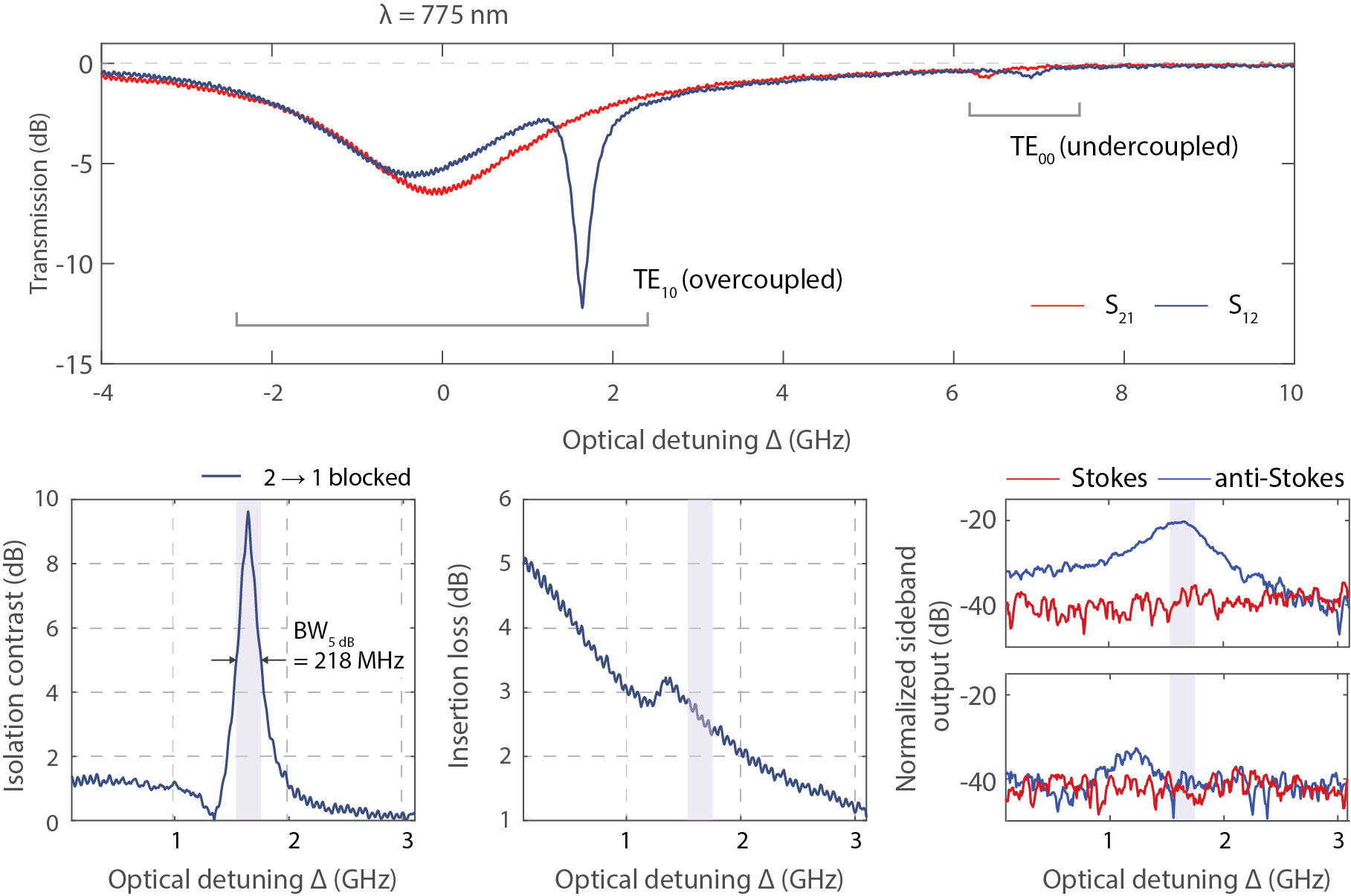}
    \caption{
        \textbf{780 nm example of giant isolation using dressed state nonreciprocity.}
        Detuning $\Delta$ is defined relative to the unperturbed TE$_{10}$ mode. 
        Here, the phase matching condition occurs in the Port 2 to Port 1 direction, since the TE$_{10}$ mode is located at lower frequency (unlike in Fig.~\ref{fig:3} and Fig.~\ref{fig:4} of the main manuscript).
        Measurement results from the 775 nm isolator for 5.05 GHz applied RF with 25 dBm power (measured $G_{ph} \approx 1.07$ GHz).
        Measured frequency mismatch $\delta$ is 1.5 GHz.
        }
    \label{sup:asym780}
    \end{adjustwidth}
\end{figure}

\FloatBarrier

\section{Isolator fabrication}
\label{sec:Fabrication}

Fabrication for both the 1550 nm and 780 nm isolator demonstrations begins with an LNOI wafer (Fig.~\ref{fig:2}b), having 500 nm thickness X-cut LN device layer, 2 $\mu$m buried oxide, and an high resistivity silicon substrate.
The racetrack resonators, waveguide and grating couplers are patterned with e-beam lithography, and the LN device layer is partially etched using argon-based inductively coupled plasma reactive ion etching (ICP-RIE).
    %
    %
The resulting ridge geometry (Fig.~\ref{fig:2}b) confines light laterally, while the underlying SiO$_2$ layer, due to its lower index, prevents leakage to the silicon substrate below.
For 780 nm isolators, we first etch back the device layer to 300 nm prior to the definition of the photonic structures, to ensure single-mode operation of the waveguide and two-mode operation of the racetrack. This operation may be varied locally to accommodate any other desired wavelength for the isolator within the transparency range of LN.
For acoustic actuation, we employ an aluminum (Al) interdigitated transducer (IDT) that is fabricated at +30 $^\circ$ from +y axis of the LN to ensure the highest electromechanical coupling~\cite{Kuznetsova:01}. 
The relative angle and pitch~\cite{Sohn18} of this IDT are calculated using finite element simulation of the acousto-optic coupling. 
All parameters for the tested devices are listed in Table~\ref{tab:DeviceParameters}.

\section{Experimental measurement setup}
\label{sec:Setup}

\begin{figure}[tb]
    \begin{adjustwidth}{-1in}{-1in}
    \centering
    \includegraphics[width=1.2\textwidth]{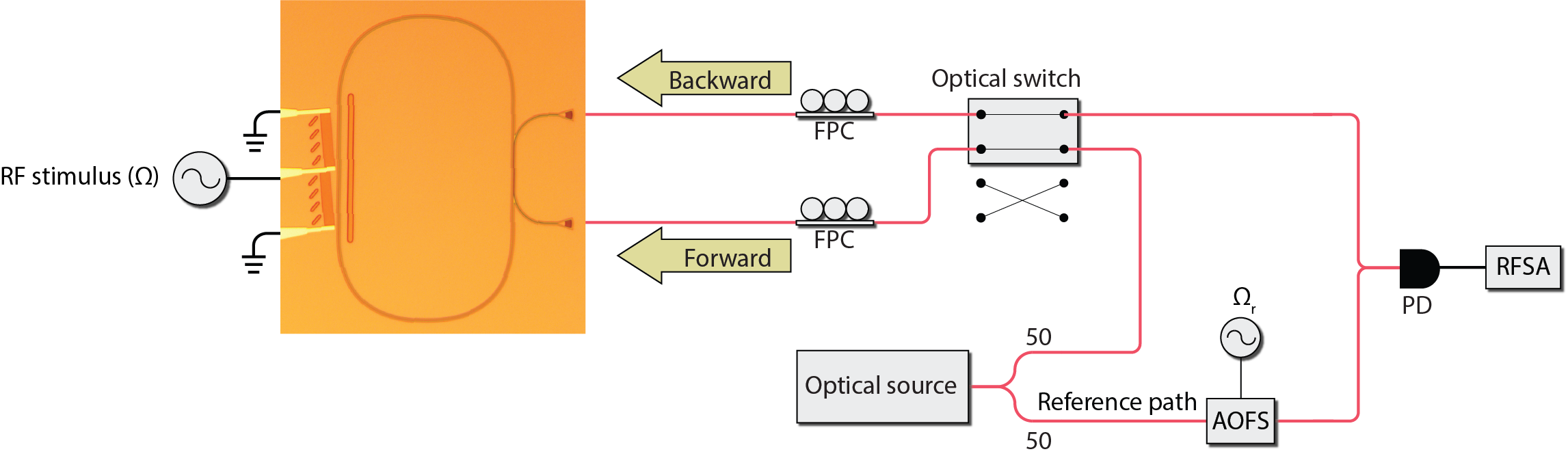}
    \caption{
        RFSA = RF Spectrum Analyzer. AOFS = Acousto-optic frequency shifter. PD = Photodetector. FPC = Fiber Polarization Controller.
    }
    \label{fig:expt_setup}
    \end{adjustwidth}
\end{figure}

For measurement of optical signal transmission, and its Stokes and anti-Stokes sidebands, we utilize an optical heterodyne detection setup as shown in Fig.~\ref{fig:expt_setup}.
The sources were $<50$ kHz linewidth tunable external cavity diode lasers; specifically the New Focus model TLB-6728-P for 1520-1570 nm, and the New Focus model TLB-6712-P for 770–783 nm.
The light is first split 50:50 into a device path and a reference path to enable heterodyne detection of
the output light spectrum. Light in the reference arm was provided a predetermined offset ($\Omega_r = $100 MHz for 1550 nm and 200 MHz for 780 nm) using an in-fibre acousto-optic frequency shifter. 
We used an optical switch (Thorlab model OSW22-1310E) to change the direction of the light for 1550 nm testing, while directionality was controlled manually for 780 nm testing.
The minimum per-grating coupler losses are $\sim 5$ dB for 1550 nm and $\sim 7$ dB for 780 nm gratings, and are factored out in our measurements. 
Temporal interference of the light from the device path and the reference path was performed by a high-frequency photodetector (Thorlabs model RXM25AF for 1550 nm and RXM25DF for 780 nm).
Further information on the experimental setup and heterodyne detection can be found in \cite{Sohn18}.

\section{Review of state of the art in on-chip optical isolators}
\label{sec:LitReview}

A comparison of monolithically integrated, non-frequency shifting, linear response isolators is presented in Table~\ref{FOM_table}. We exclude non-reciprocal modulators \cite{Sohn18, Sohn:2019aa, Kittlaus:2021wq, Kittlaus2018, sarabalis2020, Liu:19} from this comparison since the insertion loss and isolation contrast are not well defined (i.e. the definitions are architecture-dependent and author-dependent).

Magneto-optic isolators consistently exhibit larger bandwidth than both electro-optic and acousto-optic isolators.
Waveguide-based implementations consistently exhibit greater bandwidth than resonator-based isolators.

Since isolators can be cascaded, citing insertion loss (IL) or isolation contrast (IC) alone is not meaningful. Instead, the ratio of IC per dB of IL is the relevant figure of merit (FoM) for isolators~\cite{Kim2016}. We find that our best device not only provides contrast at par with the best magneto-optic isolators, but also outperforms them in insertion loss and the isolator FoM.


\newcommand{\NA}[0]{\cellcolor{black!15}N/A}
\newcommand{\AO}[0]{\cellcolor{green!15}AO}
\newcommand{\EO}[0]{\cellcolor{blue!15}EO}
\newcommand{\MO}[0]{\cellcolor{pink!15}MO}

\begin{landscape}
	\begin{centering}
		\begin{table}[h]
		\small

			\vspace{-60pt}
			\caption{Comparison of monolithically integrated, non-frequency shifting, linear response optical isolators}
			
			\begin{tabular}{| >{\raggedright\arraybackslash}m{3cm} | >{\raggedright\arraybackslash}m{0.9cm} | >{\raggedright\arraybackslash}m{1.6cm} | >{\raggedright\arraybackslash}m{1.5cm} | >{\raggedright\arraybackslash}m{1.9cm} | >{\raggedright\arraybackslash}m{1.6cm} | >{\raggedright\arraybackslash}m{1.8cm} |
			>{\raggedright\arraybackslash}m{1.8cm} | >{\raggedright\arraybackslash}m{1.8cm} | >{\raggedright\arraybackslash}m{1.6cm} |}
				
				\hline
				Author & Year & Device Type & Technique & Peak isolation contrast (IC) & Insertion loss (IL) & Peak IC per 1 dB IL (dB/dB) & 10 dB IC Bandwidth & 20 dB IC Bandwidth & Wavelength  \\ \hline \hline
				
				This work & - & Resonator & \AO & 39.31 dB & 0.65 dB & 60.46 & 108.8 MHz & 33.1 MHz & 1550 nm \\ \hline
				This work & - & Resonator & \AO & 12.75 dB & 1.13 dB & 11.28 & 198.5 MHz & \NA & 1550 nm \\ \hline 
				This work  & - & Resonator & \AO & 12.86 dB & 4.76 dB & 2.7 & 202 MHz & \NA & 780 nm\\ \hline
				\hline
				
				Kim et. al.~\cite{Kim:2021te}& 2021 & Resonator & \AO & 3 dB & 9 dB & 0.33 & \NA & \NA & 1550 nm  \\ \hline 
				\hline

                Dostart et. al.~\cite{Dostart:2021uh}& 2021 & Resonator & \EO & 13.1 dB & 18.1 dB & 0.18 & 2 GHz & \NA & 1550 nm \\ \hline 
				Doerr et. al.~\cite{Doerr:14}& 2014 &  Waveguide & \EO & 6 dB & 4 dB & 0.66 & \NA & \NA & 1550 nm \\ \hline
				 Tzuang et. al.~\cite{Tzuang2014}& 2014 & Waveguide & \EO & 2.4 dB & \NA & \NA & \NA & \NA & 1550 nm \\ \hline
			    Lira et. al.~\cite{lira2012} & 2012 & Waveguide & \EO & 3 dB & 70 dB & 0.04 & \NA & \NA & 1550 nm \\ \hline 
			    Doerr et. al.~\cite{Doerr:11}& 2011 & Waveguide & \EO & 2 dB & 11 dB & 0.18 & \NA & \NA & 1550 nm \\ \hline
			    \hline

                Yan et. al. \cite{Yan:20} & 2020 & Resonator& \MO & 28 dB & 1 dB & 28  & 6.2 GHz	$^\dagger$ & 1.2 GHz	$^\dagger$ & 1550 nm  \\ \hline
                Yan et. al.  \cite{Yan:20} & 2020 & Waveguide& \MO & 32 dB & 2.3 dB & 13.91  & 1.6 THz	$^\dagger$ & 490 GHz	$^\dagger$ & 1550 nm  \\ \hline

                Pintus et. al.~\cite{Pintus:19} & 2019 & Waveguide & \MO & 30 dB & 18 dB & 1.67  & 1.7 THz	$^\dagger$ & 582 GHz	$^\dagger$ & 1550 nm  \\ \hline
                
			    Zhang et. al.~\cite{Zhang:19} & 2019 & Waveguide & \MO & 30 dB & 5 dB & 6  & 1.1 THz	$^\dagger$ & 249.7 GHz	$^\dagger$ & 1550 nm  \\ \hline
			    Zhang et. al.~\cite{Zhang:19} & 2019 & Resonator & \MO & 20 dB & 11.5 dB & 1.739  & 1.2 GHz $^\dagger$ & \NA	$^\dagger$ & 1550 nm  \\ \hline

			    Du et. al.~\cite{Du:2018wy} & 2019 & Resonator& \MO & 40 dB & 3 dB & 13.3 & 6.2 GHz	$^\dagger$ & 3.1 GHz$^\dagger$ & 1520 nm \\ \hline
			    Bi et. al.~\cite{Ross:11}& 2019 & Resonator  & \MO & 19.5 dB & 8 dB & 2.6 & 1.6 GHz & \NA& 1520 nm \\ \hline
			    Huang et. al.~\cite{Huang:17}& 2017 & Resonator & \MO & 14.4 dB & 10 dB & 1.44 & 12.4 GHz	$^\dagger$ & \NA & 1550 nm  \\ \hline
			    Ghosh et. al.~\cite{Ghosh:12aa}& 2012 & Waveguide & \MO & 25 dB & 14 dB & 1.78 & 99 GHz	$^\dagger$ & 31 GHz	$^\dagger$ & 1490 nm \\ \hline

			\end{tabular}
			
			\vspace{12pt}
			
			AO = Acousto-optic.  EO = Electro-optic.  MO = Magneto-optic. N/A = Information not available or not applicable.
			
			$^\dagger$ Estimated from the data shown in the publication.
			

			\label{FOM_table}

		\end{table}
	\end{centering}
\end{landscape}

\begin{landscape}
	\begin{centering}
		\begin{table}[ht!]
		\small

			\vspace{-60pt}
			\caption{Parameters of our experimentally demonstrated isolators}
			
			\begin{tabular}{ | >{\raggedright\arraybackslash}m{3.5cm} |  >{\raggedright\arraybackslash}m{6cm}  >{\raggedright\arraybackslash}m{1cm}  >{\raggedright\arraybackslash}m{1.8cm}   >{\raggedright\arraybackslash}m{1.8cm}  >{\raggedright\arraybackslash}m{1.8cm} | }
				
				\hline
				 & Parameters & Unit & 1550 nm isolator in Fig.~\ref{fig:3}a & 1550 nm isolator in Fig.~\ref{fig:4}  & 780 nm isolator in Fig.~\ref{fig:3}b \\ \hline \hline
				 & $\Delta\omega$ of TE\textsubscript{00} and TE\textsubscript{10} modes ($\omega$\textsubscript{1}  - $\omega$\textsubscript{2}) & GHz & 3 & 3.72 & 5.05 \\ \cline{2-6}
				& Total loss rate of TE\textsubscript{00} mode ($\kappa$\textsubscript{1}) & GHz & 0.11 &  0.2 & 0.17 \\\cline{2-6}
				
			    & Total loss rate of TE\textsubscript{10} mode ($\kappa$\textsubscript{2}) & GHz & 1.03 & 0.98 & 1.62 \\\cline{2-6}
                & Quality factor of TE\textsubscript{00} mode (Q\textsubscript{1}) & - & 1.76 $\times$ 10\textsuperscript{6} & 0.97 $\times$ 10\textsuperscript{6} &2.25 $\times$ 10\textsuperscript{6} \\\cline{2-6}
                Two-mode optical WGR & Quality factor of TE\textsubscript{10} mode (Q\textsubscript{2}) & - & 1.88 $\times$ 10\textsuperscript{5} & 1.97 $\times$ 10\textsuperscript{5} &2.36 $\times$ 10\textsuperscript{5} \\\cline{2-6}
                & External coupling rate ($\kappa$\textsubscript{ex1}) & GHz & 0.01 & 0.04 & 0.01 \\\cline{2-6}
                & External coupling rate ($\kappa$\textsubscript{ex2}) & GHz & 0.64 & 0.81 & 0.94 \\\cline{2-6}
                & Wave number difference ( $\vert$ k\textsubscript{1} - k\textsubscript{2} $\vert$) & m\textsuperscript{-1} & 2.47 $\times$ 10\textsuperscript{5} & 2.47 $\times$ 10\textsuperscript{5} & 6.45 $\times$ 10\textsuperscript{5} \\\cline{2-6}
                & Center Wavelength & nm & 1556 & 1526 & 773 \\\cline{2-6}
                
                \hline\hline
                
                & Transverse wave number (q\textsubscript{transverse}) & m\textsuperscript{-1} & 2.84 $\times$ 10\textsuperscript{6} & 2.84 $\times$ 10\textsuperscript{6} & 5.19 $\times$ 10\textsuperscript{6} \\\cline{2-6}
                & Propagating wave number (q\textsubscript{propagating}) & m\textsuperscript{-1} & 2.47 $\times$ 10\textsuperscript{5} & 2.47 $\times$ 10\textsuperscript{5} & 6.45 $\times$ 10\textsuperscript{5} \\\cline{2-6}
                & Total wave number (q\textsubscript{total}) & m\textsuperscript{-1} & 2.85 $\times$ 10\textsuperscript{6} & 2.85 $\times$ 10\textsuperscript{6} & 5.23 $\times$ 10\textsuperscript{6} \\\cline{2-6}
                Surface Acoustic Wave & IDT Pitch ($\lambda$) & $\mu$m & 2.2 & 2.2 & 1.2 \\\cline{2-6}
                & IDT Aperture (W) & $\mu$m & 400 & 400 & 300 \\\cline{2-6}
                & IDT Angle ($\theta$) & degree  & 4.98 & 4.98 & 7.08 \\\cline{2-6}
                & Center frequency ($\Omega$) & GHz & 3 & 3.04 & 5.05 \\\cline{2-6}
                 \hline\hline
                Acousto-optic Interaction & Phonon-enhanced optomechanical coupling  & GHz & 0.97 (29 dBm) & 0.76 (29 dBm) & 0.98 (25 dBm) \\\cline{2-6}
                \hline\hline

  \end{tabular}%
   
    \label{tab:DeviceParameters}
    \vspace{24pt}
    
		\end{table}
	\end{centering}
\end{landscape}
{\footnotesize \putbib}
\end{bibunit}

\end{document}